# Generic Decoding of Seen and Imagined Objects using Hierarchical Visual Features

**Authors:**

Tomoyasu Horikawa[1] and Yukiyasu Kamitani[1,2]*

**Affiliations:**

[1]ATR Computational Neuroscience Laboratories, 2-2-2 Hikaridai, Seika, Soraku, Kyoto 619-0288, Japan

[2]Graduate School of Informatics, Kyoto University, Yoshida-honmachi, Sakyo-ku, Kyoto 606-8501, Japan

*Corresponding author's email address: kamitani@i.kyoto-u.ac.jp

# Abstract

Object recognition is a key function in both human and machine vision. While recent studies have achieved fMRI decoding of seen and imagined contents, the prediction is limited to training examples. We present a decoding approach for arbitrary objects, using the machine vision principle that an object category is represented by a set of features rendered invariant through hierarchical processing. We show that visual features including those from a convolutional neural network can be predicted from fMRI patterns and that greater accuracy is achieved for low/high-level features with lower/higher-level visual areas, respectively. Predicted features are used to identify seen/imagined object categories (extending beyond decoder training) from a set of computed features for numerous object images. Furthermore, the decoding of imagined objects reveals progressive recruitment of higher to lower visual representations. Our results demonstrate a homology between human and machine vision and its utility for brain-based information retrieval.

# Introduction

Brain decoding through machine learning analysis of functional magnetic resonance imaging (fMRI) activity has enabled the interpretation of mental contents, including what people see[1,2], remember[3-7], imagine[8-12], and dream[13]. Most of the studies relied on a classification-based approach, where a statistical classifier (decoder) is trained to learn a relationship between fMRI patterns and the target contents to be decoded. Such an approach, however, entails a fundamental constraint on the number of possible outputs. Namely, the outputs are limited to the classes used for decoder training, preventing the decoder from predicting any classes that are not used in training.

Recent studies have overcome the limitation on the number of possible outputs by designing decoders for retinotopically organized, image-level features[14-16]. This enabled the decoding of novel visual images that were not presented during training sessions. Kay et al. (2008)[14] built an encoding model consisting of retinotopically organized Gabor wavelet filters. They used a database of visual images and the predicted brain activities produced by an encoding model. The measured brain activity was then decoded by determining the visual image in the database corresponding to the predicted brain activity most similar to the measured brain activity. This technique can also be used to identify remembered artworks from the early visual cortical activity patterns[17]. Miyawaki et al. (2008)[16] constructed a modular decoding model consisting of multi-scale local decoders (modules) that predicted the contrast of local image patches of various scales. The model enabled reconstruction of arbitrary visual images from brain activity by combining the outputs of the local decoders despite having been trained with brain activities for a small number of random images.

While the visual image identification[14,15,17] and reconstruction[16] are suitable for decoding according to image-based similarity, they do not provide explicit information regarding the object a person is seeing or imagining. The possible objects we may see or imagine in daily life are countless, and object-based information is often of more direct relevance to our visually guided behavior than image-based information. Establishing a method to decode generic object categories from brain activity would provide practical benefits for technologies utilizing information decoded from brain activity, and may enrich our understanding of how the human brain represents a vast number of objects.

In this study, we aim to decode seen and imagined object categories, including those not used in decoder training, from fMRI signals measured while subjects either viewed or imagined object images. We extend the modular decoding approach originally developed for visual image reconstruction[16] to the decoding of generic object categories.

To tailor the modular decoding approach to our objectives, we assumed that an object category can be represented by a set of visual features with several invariances. These features correspond to those proposed for the object recognition challenge in machine vision[18-24] (Fig. 1a), which aims at enabling a computer to recognize objects in images according to their category names. The selection of the visual features is a critical aspect of this approach because even if images depict the same object, they do not necessarily have pixel-wise similarity as a result of varying rotation, scale, position, and other attributes. Thus, objects may be more suitably represented using mid- or high-level visual features which are invariant to such image differences rather than the low-level features (e.g., local contrast[13] or Gabor wavelet filter[14,15,17]) used for visual image reconstruction and image identification.

We tested a total of 13 candidates of visual feature types/layers constructed from four models (Fig. 1a, see Methods): a convolutional neural network (CNN) model[20] (CNN1-CNN8), HMAX model[21-23] (HMAX1-HMAX3), GIST[24], and scale invariant feature transform[18] combined with "Bag of Features"[19] (SIFT+BoF). Some of the models emulate the structure of hierarchical human visual system (CNN and HMAX), and the others are models designed for scene recognition (GIST) and object recognition (SIFT+BoF) in machine vision. These visual feature types/layers have multiple levels of complexity, and it has been reported that representations of these model outputs are statistically similar to visual cortical activity[21,22,25-30].

Using these visual features, we present an approach called *generic object decoding* in which arbitrary object categories are decoded from human brain activity (Fig. 1b). We used the online image database, *ImageNet*[31] and trained regression models (decoders) to predict visual features extracted by the computational models from brain activities recorded by fMRI while subjects viewed natural images (150 categories). The trained decoders were then used to predict feature vectors of seen and imagined objects that were not used in decoder training from the fMRI activity patterns. By comparing the predicted feature vector with the category-average feature vectors calculated from images in the image database, we identify seen and imagined object categories from those defined in the database (15,372 categories in *ImageNet*[31]). Note that because arbitrary object categories are represented in this feature space, the identified categories are not limited to those used in the training session.

Here, we first demonstrate that visual feature values of seen objects calculated by the computational models can be predicted from multiple brain areas, showing tight associations between hierarchical visual cortical areas and the complexity levels of visual features. We also show that the stimulus-trained decoders can be used to decode visual features of imagined objects, providing evidence for the progressive recruitment of hierarchical neural representations in a top-to-bottom manner. Finally, we test whether the features predicted from brain activity patterns are useful for identifying seen and imagined objects for arbitrary categories.

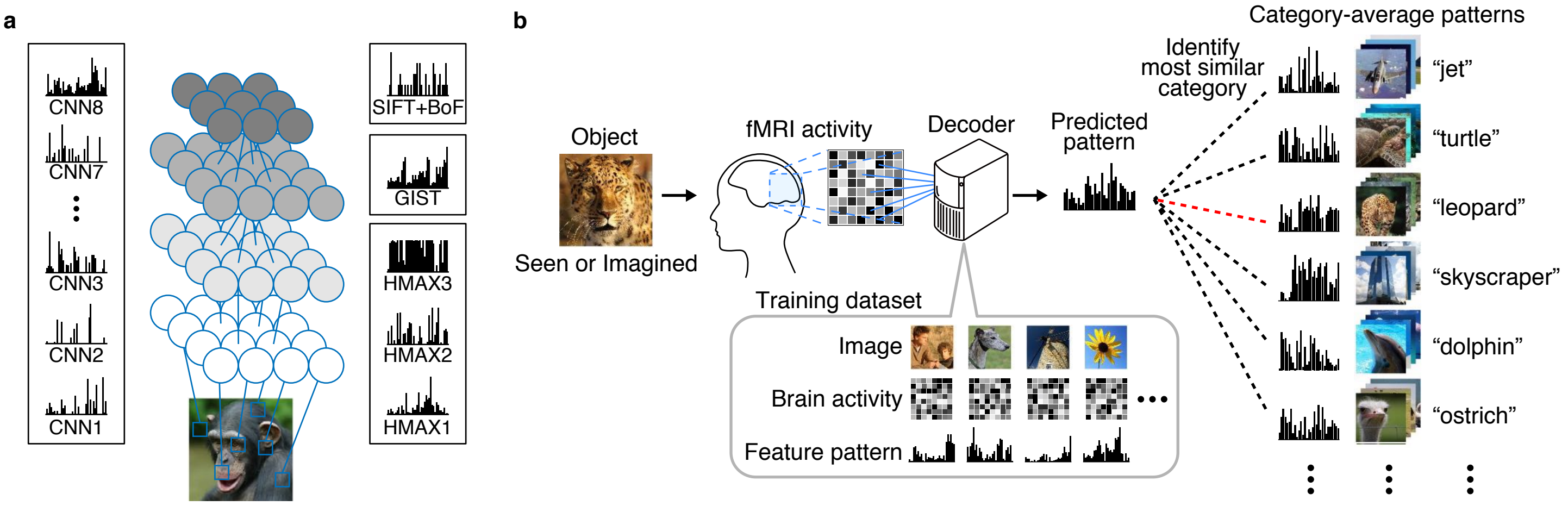

**Figure 1 | Generic object decoding.** (**a**) Visual feature extraction from natural images using computational models. Visual features are calculated from natural images using CNN (CNN1–8), HMAX (HMAX1–3), GIST, and SIFT+BoF. (**b**) Overview of generic object decoding. fMRI activity was measured while subjects viewed natural images. Decoders are trained to predict the values of the visual features for presented images/objects from multi-voxel fMRI signals. Given measured fMRI activity, a feature vector is predicted and it is used to identify the seen or imagined object by comparing it with the feature vectors of numerous objects in an annotated image database including those not used for decoder training.

# Results

### Generic object decoding

Our objective was to decode arbitrary object categories (which were not included in model training) from human brain activity measured using fMRI. Our approach consisted of the following steps. First, we extracted feature values from object images using a total of 13 visual feature types/layers from four computational models (CNN1–8, HMAX1–3, GIST, and SIFT+BoF; ~1,000 units for each feature type/layer). We thereby represented an object image by a feature vector of each feature type/layer (Fig. 1a). Second, decoders were trained to predict the vectors of visual features of seen objects from brain activity patterns (Fig. 1b). Third, using the trained decoders, a feature vector was predicted from brain activity measured while seeing or imagining an object, which was not used in decoder training. Finally, the predicted feature vector was used to identify a seen or imagined object by calculating the similarity between the predicted and the category-average feature vectors calculated from an annotated image database.

To test the feasibility of generic decoding of seen and imagined objects from human brain activities, we conducted two fMRI experiments: an image presentation experiment, and an imagery experiment (Fig. 2). In the image presentation experiment, fMRI signals were measured while subjects viewed a sequence of object images (Fig. 2a). The image presentation experiment consisted of two sessions: the training image session and test image session. In the training image session, a total of 1,200 images from 150 different object categories (eight images from each category) were each presented once. In the test image session, a total of 50 images from 50 object categories (one image from each category) were each presented 35 times. In the imagery experiment, fMRI signals were measured while the subjects imagined about one of the 50 object categories (10 times for each category) that were the same with those used in the test image session (Fig. 2b). Note that the categories in the test image session and the imagery experiment were not used in the training image session. While we show results with fMRI signals averaged across all trials (35 trials for the test image session, and 10 trials for the imagery experiment), quantitatively similar results were obtained with a much smaller number of averaged samples (see Supplementary Information).

We performed our analysis for each combination of feature types/layers (CNN1–8, HMAX1–3, GIST, and SIFT+BoF) and brain regions of interest (ROI; V1, V2, V3, V4, the lateral occipital complex [LOC], fusiform face area [FFA], parahippocampal place area [PPA], lower visual cortex [LVC; V1-V3], higher visual cortex [HVC; covering regions around LOC, FFA, and PPA], and an entire visual cortex [VC; covering all of the visual subareas listed above]; see Methods, and Supplementary Fig. 1 for the definitions of regions of interest).

A set of linear regression functions (sparse linear regression model[32]) was used to predict visual feature vectors (~1,000 feature units for each feature type/layer; see Methods) from the fMRI signals in each ROI. A unit decoder was trained to predict the values of the feature vectors calculated from the viewed images, using the fMRI signals from the training image session (*i.e.*, ~1,000 decoders for each feature type/layer). The trained decoders were then used to predict the vectors of each feature type/layer for test object categories from measured fMRI signals in the test image session and the imagery experiment.

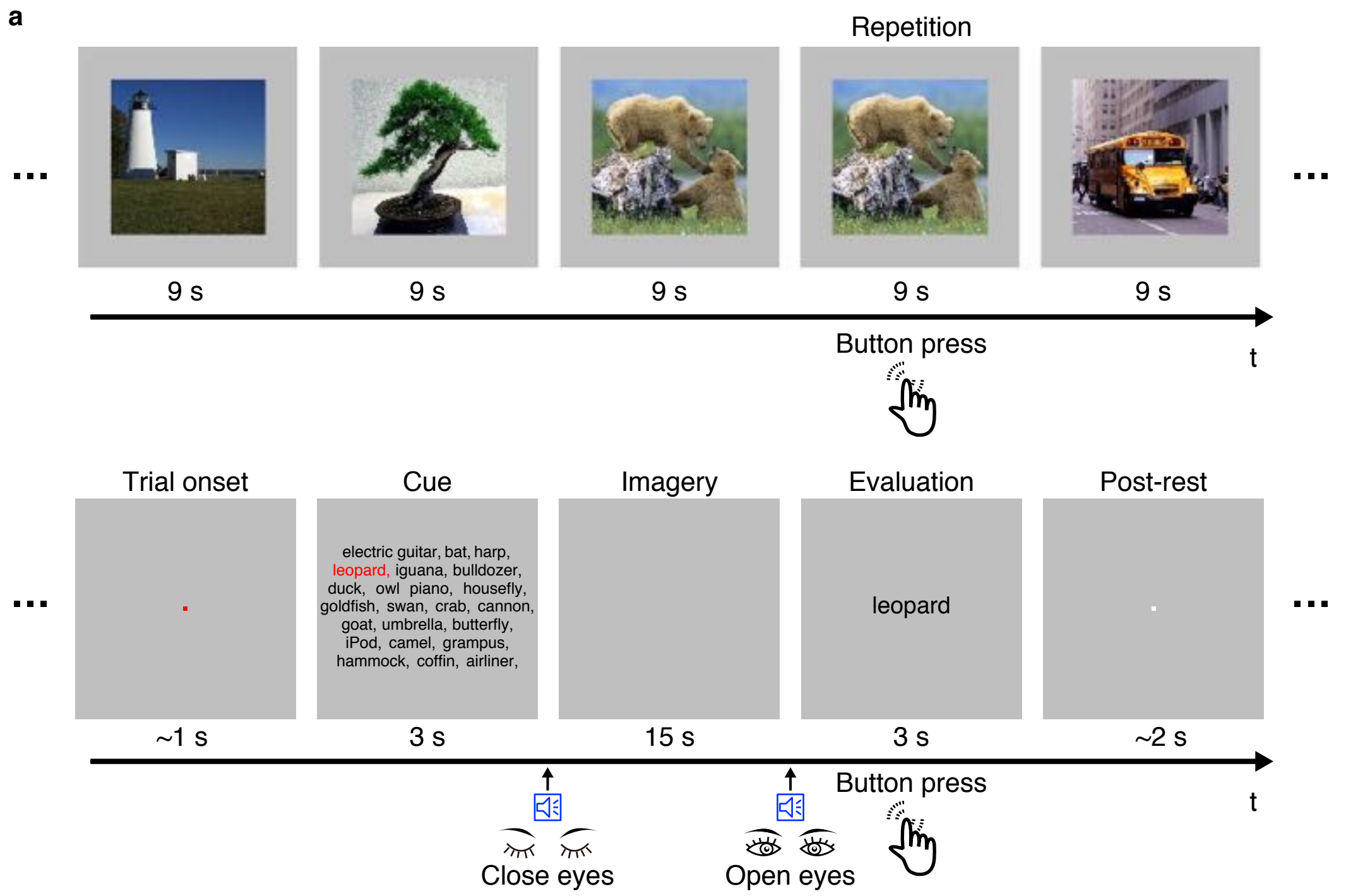

**Figure 2 | Experimental design.** (**a**) Image presentation experiment. Images were presented in the center of the display with a central fixation spot. The color of the fixation spot changed from white to red for 0.5 seconds before each stimulus block started to indicate the onset of the block. Subjects maintained steady fixation throughout each run and performed a one-back repetition detection task on the images, responding with a button press for each repetition. (**b**) Imagery experiment. The onset of each trial was marked by a change in the fixation color. Cue stimuli composed of an array of object names were visually presented for 3 seconds. The onset and the end of the imagery periods were signaled by aural beeps. After the first beep, the subjects were required to imagine as many object images as possible pertaining to the category indicated by red letters. They continued imagining with their eyes closed (15 seconds) pending a second beep. Subjects were then asked to evaluate the vividness of their mental imagery (3 seconds). Note that

the actual cue consisted of an array of 50 object names, while only subsets of the words are depicted in this figure because of space limitation.

**Image feature decoding**

We first investigated whether we could decode the values of visual feature vectors for presented images from brain activity. Decoding accuracy was evaluated by the correlation coefficient between true and predicted feature values of each feature unit (Fig. 3a). Correlation coefficients were averaged across the units in each feature type/layer for multiple ROIs, and then averaged across five subjects. Because the distribution of feature values and the number of feature units of the original population differed between feature types/layers, interpreting decoding accuracy differences across feature types/layers is difficult. For example, when the accuracy was evaluated by normalized root mean square errors, the overall pattern of accuracy across feature types/layers was different from that of correlation (Supplementary Fig. 2). Therefore, in the following we mainly focused on the pattern of accuracies across ROIs in each feature type/layer.

Fig. 3b shows the decoding accuracy for features of presented images in multiple ROIs. The predicted feature values positively correlated with the true values for all feature–ROI combinations (one-sided $t$ test after Fisher's Z transform, uncorrected $P < 0.05$). Interestingly, the choice of feature types/layers and ROIs affected the accuracy pattern.

As observed in the results for the CNN layers, higher-order features tended to be better predicted from fMRI signals in higher rather than lower ROIs, and lower-order features tended to be better predicted from fMRI signals in lower rather than higher ROIs (ANOVA, interaction between layer and ROI, $P < 0.01$). Similar tendencies were also observed in HMAX (ANOVA, interaction between feature type and ROI, $P < 0.01$). Such differences were also observed in the decoding accuracies obtained from LVC and HVC (Supplementary Fig. 3). These results reveal a tight association between hierarchical visual areas and the complexity levels of visual features in image feature decoding accuracy.

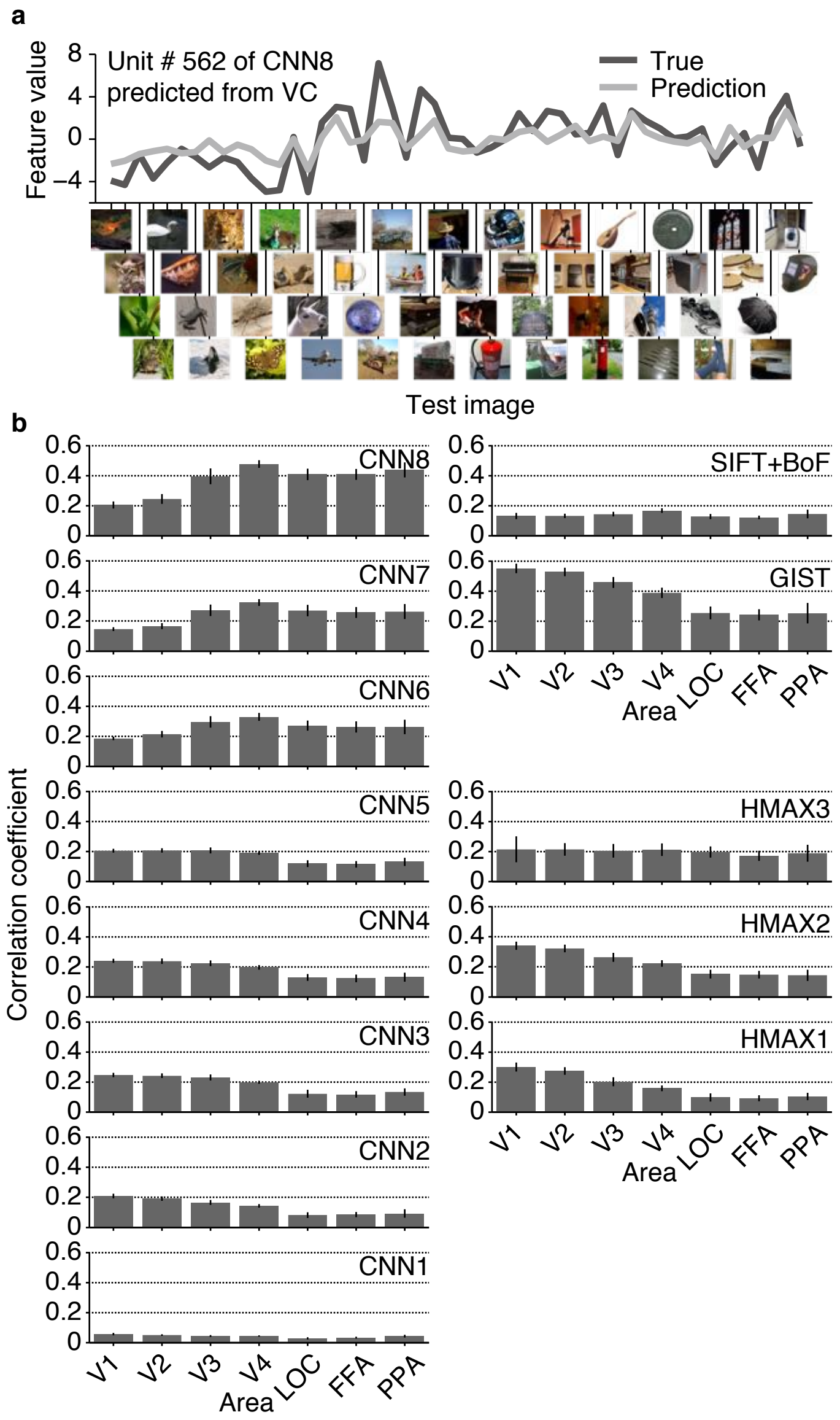

**Figure 3 | Image feature decoding and the homology of CNN and the human brain.** (**a**) Example of a feature unit. True (black) and predicted (light gray) feature values for 50 test images are shown for Unit #562 of CNN Layer 8 (CNN8) predicted from the whole visual cortex (VC). (**b**) Image feature decoding accuracies. Mean decoding accuracies are shown for each combination of the feature type/layer and ROI (error bars, 95% confidence interval (CI) across five subjects).

To understand more details about what each visual feature represents, we synthesized preferred images that highly activate individual units in each CNN layer using the activation maximization technique[33–36] (Fig. 4; see Supplementary Fig.4 for more examples; see Methods). The generated images showed a gradual increase in complexity, starting from simple edge-detector-like representations to more complex shapes, textures, or object parts, and to intact objects. Because the CNN6–8 are fully-connected layers, position information about their preferred pattern was lost. These preferred images of individual CNN units resemble the critical features found in monkey electrophysiological studies[37].

Analyses of voxel weights learned by the image feature decoders (trained with VC) showed differences in the spatial distribution of predictive voxels between visual feature types/layers. We used a linear regression model in which voxel weights were estimated with sparseness priors[32], resulting in a small subset of voxels selected with non-zero weights (hence relevant for decoding). Examples of selected sets of voxels for a CNN2 unit and a CNN8 unit are shown in Fig. 4b. While the voxels selected for predicting a CNN2 unit distributed mainly around the lower visual areas (V1–V3), those for predicting a CNN8 unit distributed around more anterior areas (Fig. 4b). The distributions of predictive voxels for the visual feature types/layers were consistent to the results of the image feature decoding analysis using individual ROIs (Fig. 3b): voxels in the lower/higher ROIs were more frequently selected for predicting lower/higher visual features, respectively (Fig. 4c; ANOVA, interaction between feature type/layer and ROI, $P < 0.01$; see Supplementary Fig. 5 for distributions for HMAX1–3, GIST, and SIFT+BoF).

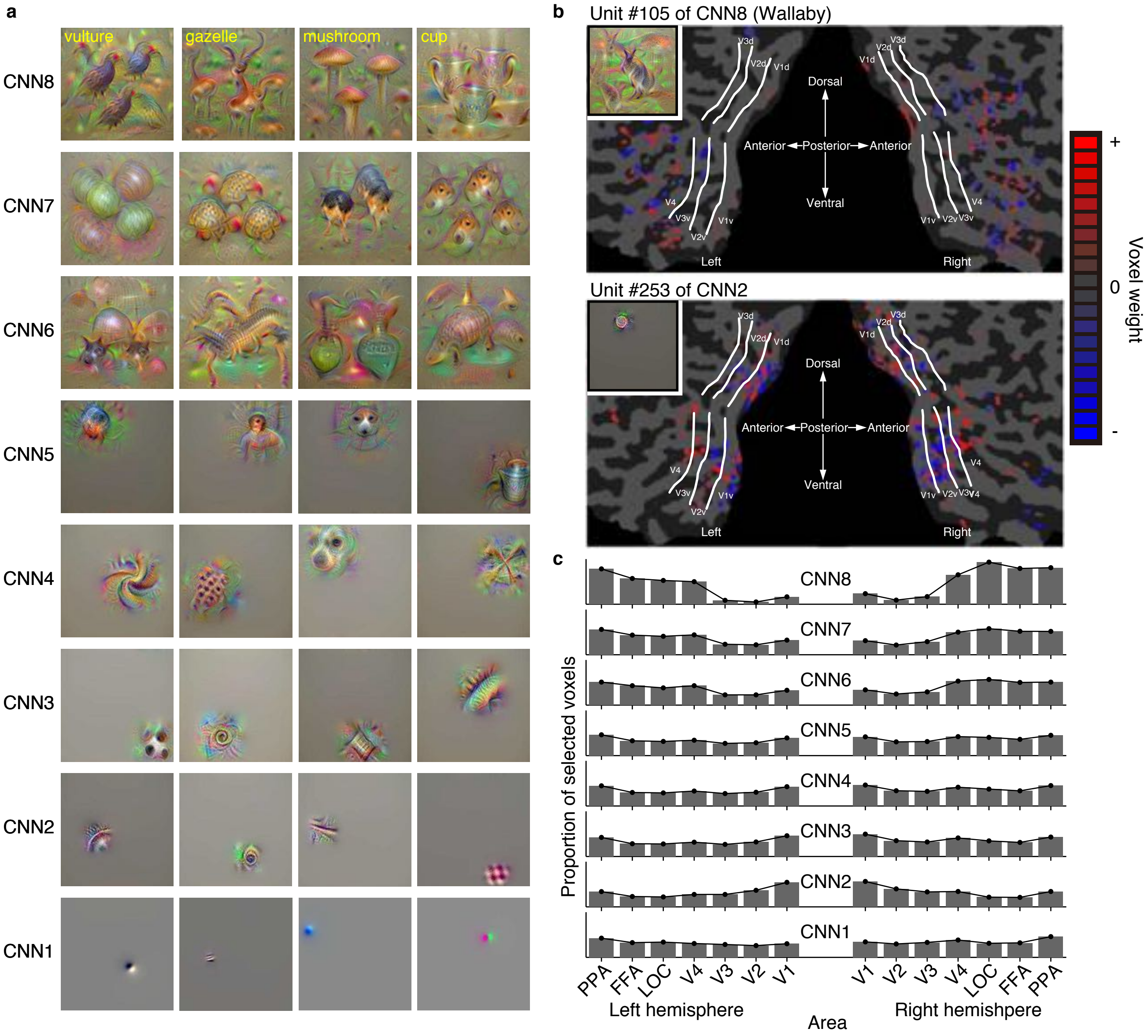

**Figure 4 | Preferred images and weight distributions for CNN layers.** (**a**) Examples of preferred images for individual units (four randomly selected units) in each CNN layer. Because each unit of CNN8 corresponds to a specific category, the names of the categories for each unit are shown on the left top. (**b**) Voxel weight maps on the flattened cortical surface. Weights resulting from the decoders trained to predict feature values of a single feature unit are shown for

Unit #105 of CNN8 and for Unit #253 of CNN2 (predicted from VC, Subject 3). The preferred image for the unit is indicated in the inset. (**c**) Distributions of selected voxels across individual subareas for the CNN layers. Distributions of selected voxels used for prediction are shown for each CNN layer (predicted from VC, five subjects averaged). The proportion of selected voxels for each subarea was calculated by first counting the numbers of selected voxels for individual feature units, aggregating them over ~1,000 feature units in each layer, and then normalizing with the total voxels.

Additionally, we found that the image feature decoding accuracy in each unit was positively correlated with the “category discriminability” of each unit (Fig. 5). As an index of category discriminability, we calculated the ***F*** statistic of each feature unit (a ratio of inter- and intra-category variations of feature values calculated from images in the *ImageNet* [15,322 categories]). The distributions of the category discriminability from each feature type/layer showed that the high-level features tend to demonstrate high discriminability for both CNN and HMAX (Fig. 5a bottom). Within each feature type/layer, positive correlations between decoding accuracy and category discriminability were observed for all visual features except for HMAX1 (Fig. 5b and c). Furthermore, decoding accuracy and category discriminability were positively correlated even when feature units were combined across all feature types/layers (Fig. 5c, All) and were averaged within each feature type/layer (Fig. 5c, Mean). While this analysis was performed with decoders trained on the whole visual cortical activity (VC), this tendency was robustly reproduced with each ROI. Thus, decodable feature units tend to be critical for defining object categories.

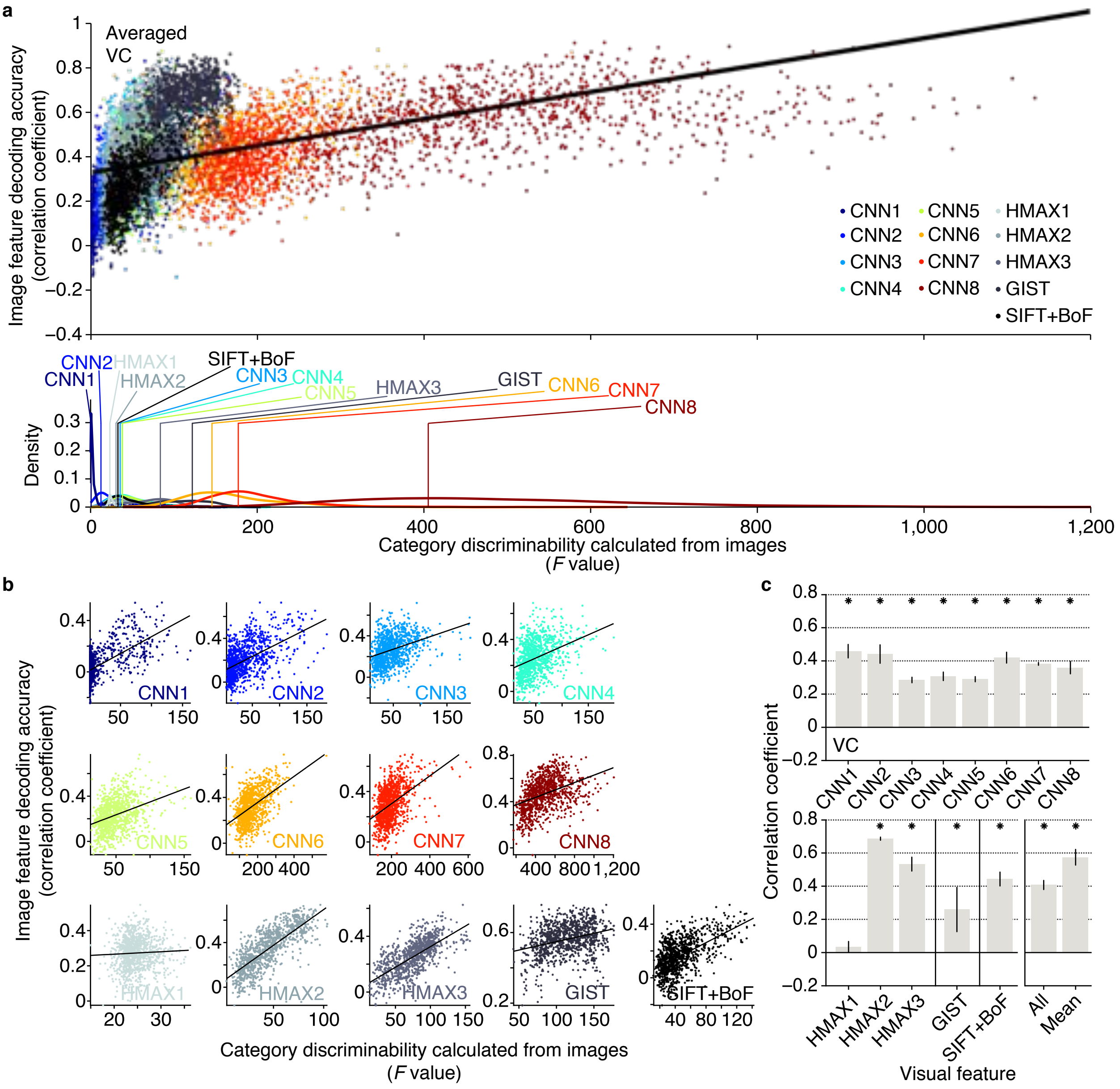

**Figure 5 | Category discriminability versus decodability of individual feature units. (a)** Scatterplot of image feature decoding accuracy against category discriminability for all the feature types/layers (top panel, predicted from VC, average over five subjects for each unit), and the distributions of category discriminability for each feature type/layer (bottom panel; vertical lines denote the mode of each distribution). Each dot in the scatter plot denotes the *F* statistic and the

decoding accuracy for a feature unit (~1,000 units). (**b**) Scatterplots for individual feature types/layers. (**c**) Correlation coefficients between category discriminability and image feature decoding accuracy (error bars, 95% CI across five subjects; asterisks, one-sided *t* test after Fisher's Z transform, uncorrected $P < 0.05$). The solid line in the scatterplots indicates a fitted regression line.

**Prediction of category-average features from stimulus- and imagery-induced brain activity**

In computer vision, objet recognition is often performed by matching the feature vector of an input image with a set of category-specific feature vectors, assuming that hierarchical features are progressively rendered invariant to represent object categories. To link the image feature decoding with object recognition, we next tested whether the feature values from the image feature decoders (cf., Fig. 3) predicted the values of category-specific feature vectors. We constructed category-specific feature vectors by averaging the feature vectors of multiple images annotated with the same object category (15,372 categories in *ImageNet*[31]). To evaluate the prediction accuracy in each unit, Pearson's correlation coefficient was calculated between the predicted and the category-average feature values for the series of test trials. We then averaged the correlation coefficients across the units in each feature type/layer. The evaluation with category-average features allowed us to extend the feature decoding analysis to the imagery experiment, in which subjects freely imagined about an object cued by text, and thus there were no ground truth images from which visual features could be calculated.

The correlation coefficients between the features decoded from stimulus- and imagery-induced brain activity and the category-average features in multiple ROIs are shown in Fig. 6 (see Supplementary Fig. 8 for distributions of correlation coefficient for individual units). The features decoded from stimulus-induced brain activity were significantly correlated with the category-average features for all feature–ROI combinations (Fig. 6a; one-sided *t* test after Fisher's Z transform, uncorrected $P < 0.05$). In contrast to image features (Fig. 3b), category-average features were better predicted in higher than lower ROIs for most feature types/layers. Imagery-induced brain activity showed a similar pattern of prediction but with reduced accuracies: relatively high correlations were found for mid-to-high level CNN features using mid-to-high level ROIs (V4, LOC, FFA, and PPA) (Fig. 6b). Thus, image features decoded from stimulus- and imagery-induced brain activity are predictive of category-specific features, especially with mid-to-high level CNN features decoded from mid-to-high level ROIs. Poorer prediction with other features and ROIs may be due to the lack of invariance to image attributes irrelevant for object recognition. The capacity of imagery-induced brain activity to predict mid-level, as well as top-level CNN features suggests that mental imagery may recruit neural representations of visual features with intermediate complexity, which are not simply pictorial or conceptual, via progressive top-down processing.

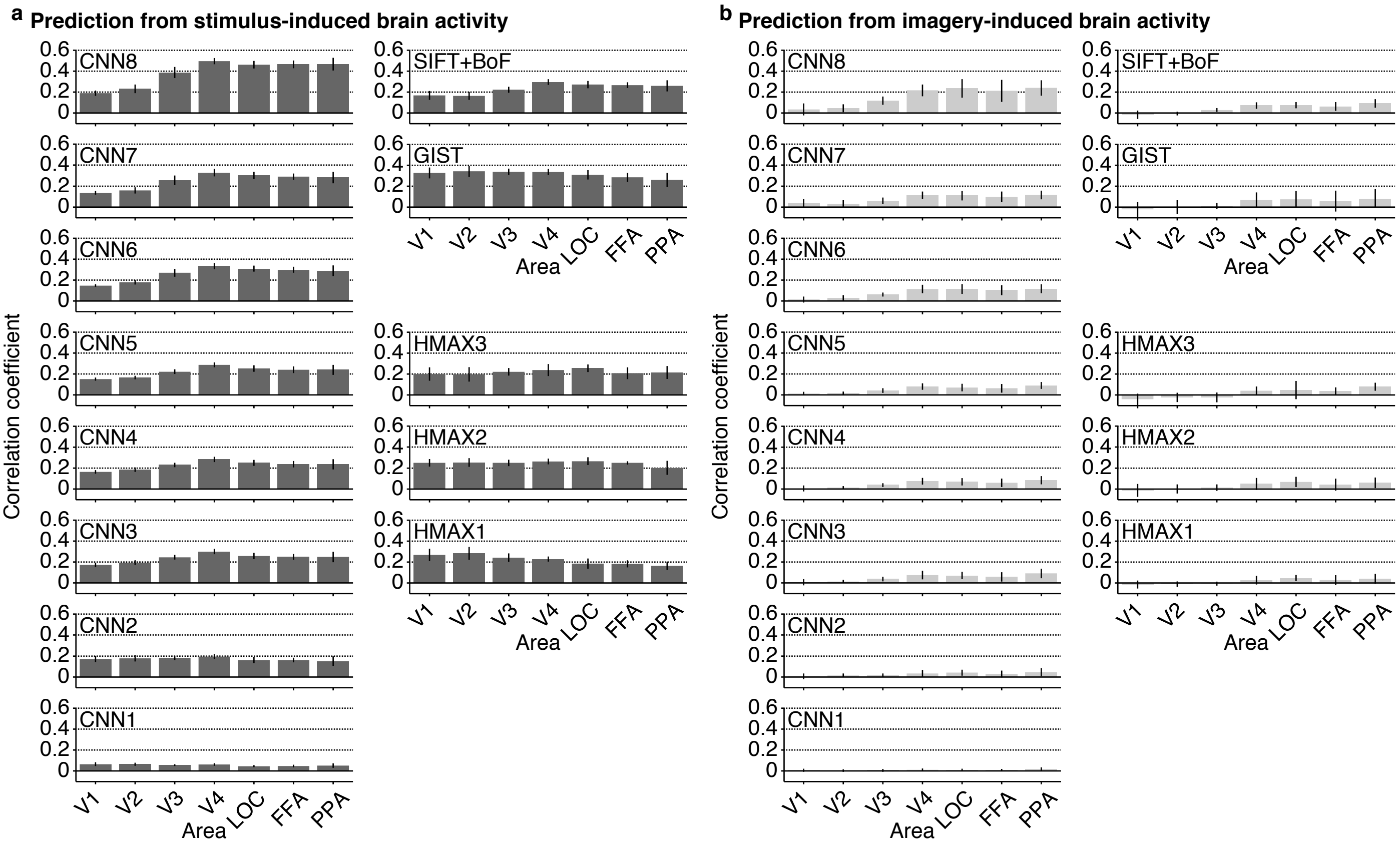

**Figure 6 | Prediction of category-average features from stimulus- and imagery-induced brain activity**. (**a**) Correlation coefficients with predicted features from stimulus-induced brain activity. (**b**) Correlation coefficients with predicted features from imagery-induced brain activity. Mean correlation coefficients are shown for each feature type/layer and ROI (error bars, 95% CI across five subjects). See Supplementary Fig. 6 for the relation between category discriminability and prediction accuracy (cf., Fig. 5c). Similar analyses can be performed using the decoders trained with category-average features (not image features) for training stimulus images (category-average feature decoders), showing qualitatively similar prediction results with higher accuracies with the imagery-induced brain activity (Supplementary Fig. 7 and 8b).

To find out a signature of such progressive top-down processing, we performed a time-resolved feature prediction analysis. While the results in Fig. 6 are based on the average brain activity during the entire 9-s stimulus or 15-s imagery period, the time course of prediction accuracies at each time point reveals differences between CNN layers and ROIs (Fig. 7). When the features in each CNN layer were predicted from imagery-induced brain activity in the whole visual cortex (VC), the peak timings for higher CNN layers tended to precede those for lower CNN layers, except CNN1, which shows poor prediction and no clear peak (Fig. 7a, b; ANOVA, interaction between time and CNN layer, $P < 0.01$). Similarly, when feature prediction accuracies for all layers were averaged for each ROI, the peak timings for higher ROIs tended to precede those for lower ROIs (Fig. 7c, d; ANOVA, interaction between time and ROI, $P < 0.01$; see Supplementary Fig. 9 for time courses for each CNN layer). Such time differences across CNN layers or ROIs were not found with stimulus-induced brain activity (Fig. 7e, f, g, and h). Although anecdotally, subjects reported that in the imagery task of this study, it often took several seconds for vivid visual imagery to develop. The imagery task may have progressively activated hierarchical neural representations in a top-to-bottom manner over several seconds in concert with the vividness of imagery.

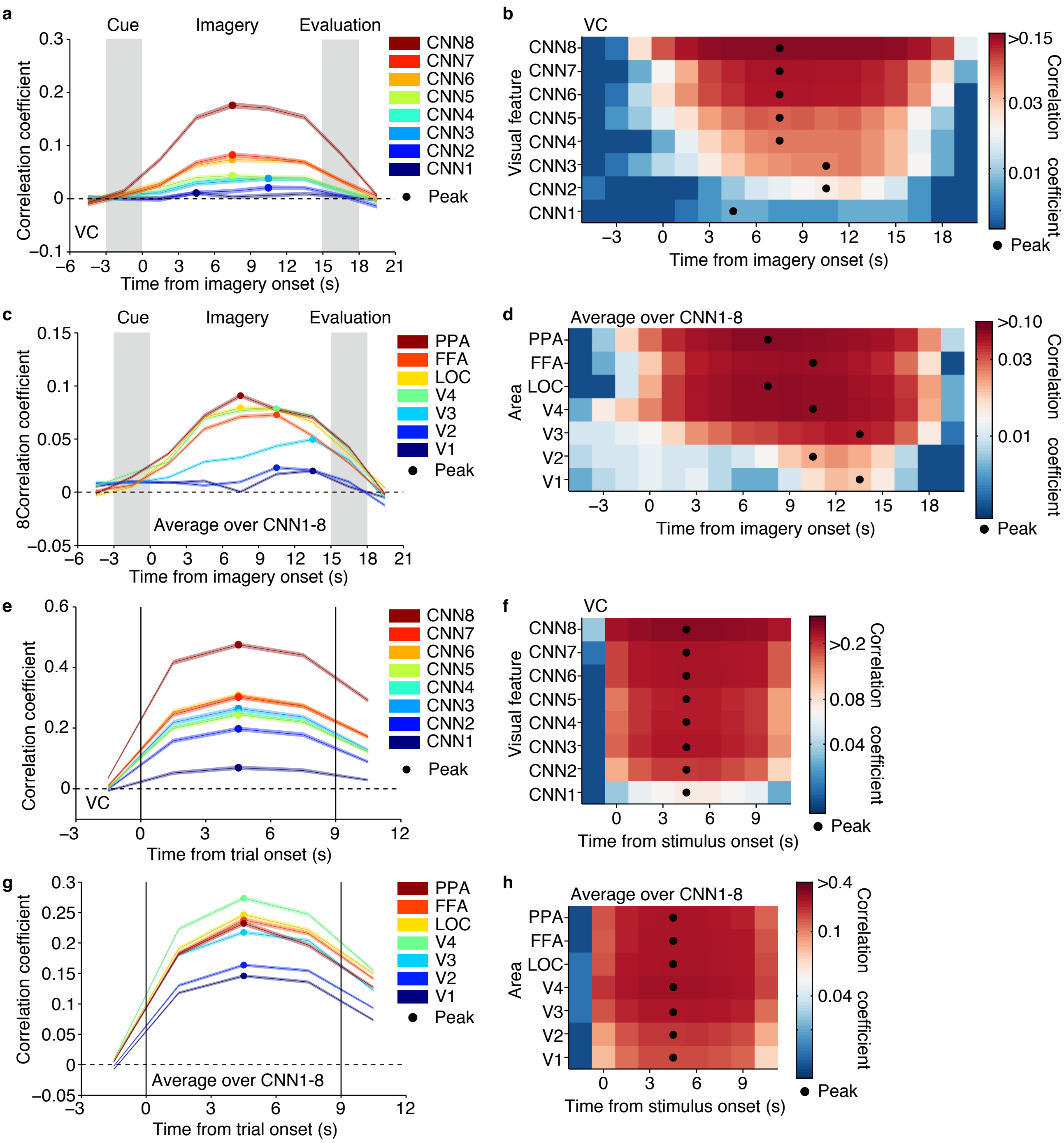

**Figure 7 | Time course of feature prediction from imagery- and stimulus-induced brain activity**. At each time point/volume around the task period, correlation coefficients were calculated between the predicted and the category-average feature values for the series of test trials (five subject averaged; shaded areas, 95% CI across feature units; filled circles, peak timing). (**a, b**) Line plots and color display for prediction from imagery-induced brain activity (VC) at

different CNN layers. (**c, d**) Prediction from imagery-induced brain activity in individual ROIs (average over CNN1–8). (**e, f**) Prediction from stimulus-induced brain activity (VC) at different CNN layers. (**g, h**) Prediction from stimulus-induced brain activity in individual ROIs (average over CNN1–8). Analyses were performed on the volume-by-volume basis while colormaps are drawn by averaging time courses evaluated by every single volume or every 2 or 3 volumes average by allowing overlap for display purposes.

## Object category identification

We next conducted identification analysis[14,16] to examine whether a predicted feature vector is useful for identifying the seen or imagined object. Because our approach is not constrained by the categories used for decoder training, we can perform identification analysis for thousands of object categories, including those not used for model training. Here, the category of the seen or imagined object was identified from a variable number of candidate categories (Fig. 8). We constructed the candidate feature vector set consisting of object categories used in the test image session (and the imagery experiments) and a specified number of object categories randomly selected from the 15,322 categories provided by *ImageNet*[31]. Given an fMRI sample, category identification was performed by selecting the category-average feature vector with the highest correlation coefficient with the predicted feature vector.

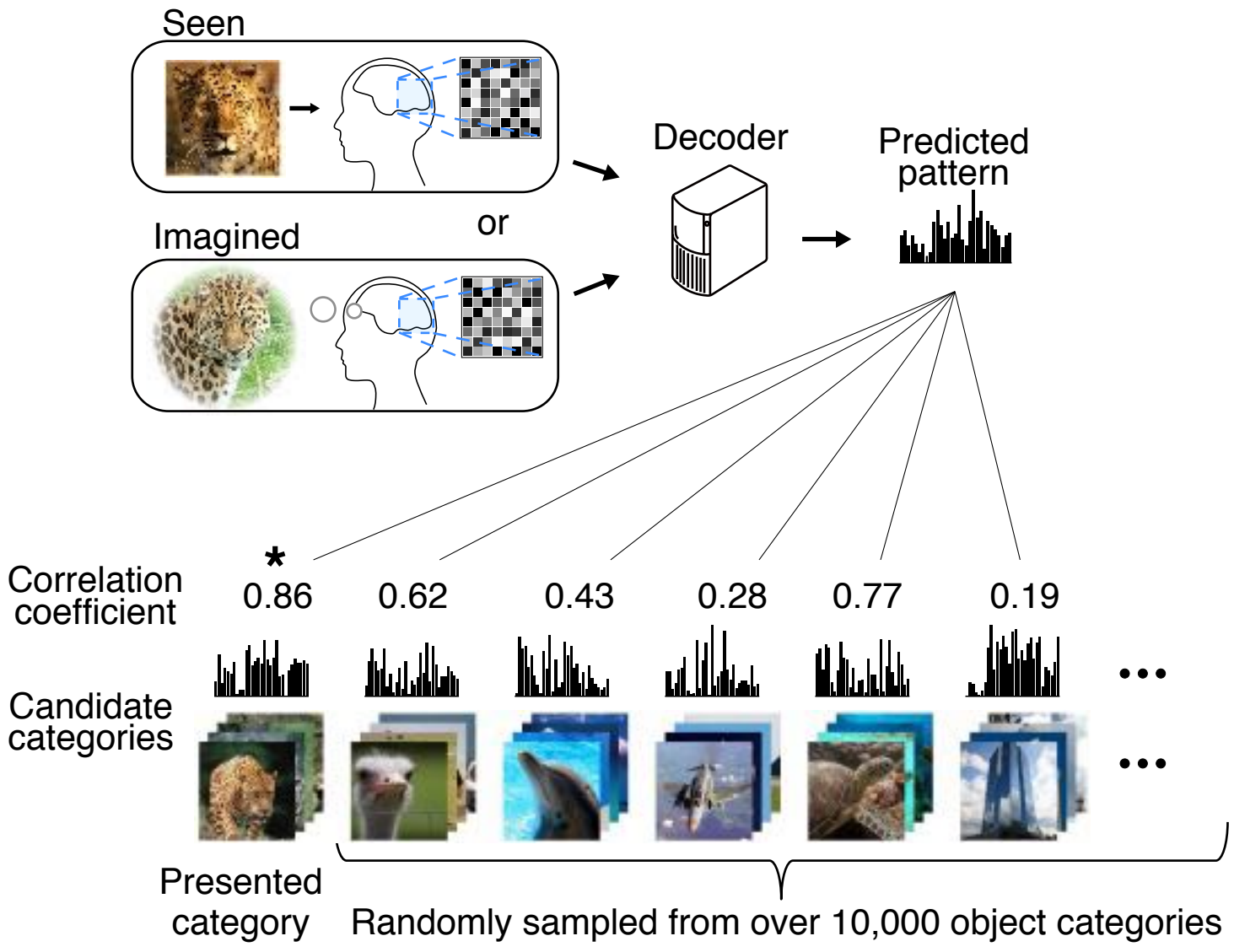

**Figure 8 | Category identification procedure.** The correlation coefficient was calculated between a predicted feature vector and the category-average feature vectors for categories in the candidate set, consisting of the presented or imagined category and a specified number of categories randomly selected from the *ImageNet* database[31]. The category with the highest correlation coefficient was selected as the predicted category (marked by a star).

First, we illustrate examples with the top six categories selected from 1,000 candidates (Fig. 9a). For both seen and imagined objects, the true categories were correctly selected or highly ranked. Even when the correct categories were not assigned, the top six categories appeared to include similar or related categories (e.g., the decoded feature vector for "duck" misidentified another type of bird "solitaire").

Next, we quantitatively evaluated the relations between rank and semantic distance with respect to the target categories and the categories ranked in each position. The semantic distance was defined by the shortest path length between the categories in the *WordNet* tree[38], and was calculated between the target category and each of the 1,000 candidate categories. The 1,000 distances were then sorted by the object category ranking (the similarity to the decoded feature vector), and averaged over 1,000 repetitions of random candidate selection and 50 target categories. The analysis showed that the categories ranked in higher positions tended to show shorter semantic distance to the target categories (Fig. 9b). The semantic distance was positively correlated with rank, especially for mid-to-high level CNN layers (CNN3–8) and SIFT+BoF under both seen and imagined conditions (Fig. 9c; asterisks, one-sided $t$ test after Fisher's Z transform, uncorrected $P < 0.05$). These results suggest that for these feature types/layers, semantically similar, if not correct, categories can be selected with the feature vector predicted from brain activity.

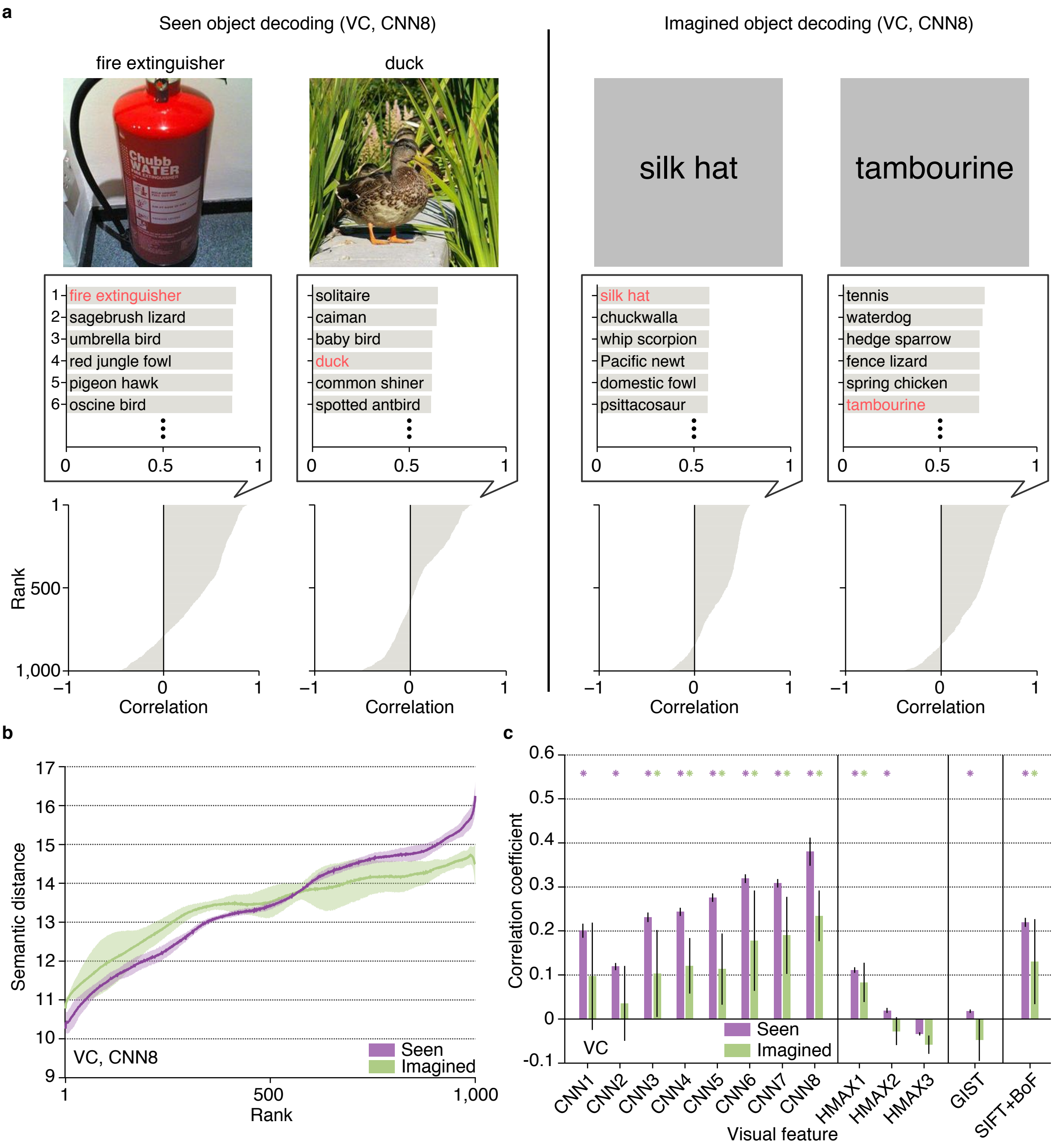

**Figure 9 | Object categories ranked by the similarity to decoded feature vectors. (a)** Object category rankings and correlation scores. Rankings of object categories are shown for two viewed objects ("fire extinguisher" and "duck") and two imagined objects ("silk hat" and "tambourine",) (CNN8, predicted from VC; candidate sets consisting of each one of the true categories [$n = 50$]

and 999 randomly selected false categories). The candidate object names and correlation scores for the top six objects with the highest correlation coefficients are shown in the middle panels (correct object names indicated in red). The panels at the bottom illustrate the correlation coefficients for 1,000 object categories in descending order. (**b**) Relationship between the object category ranking and the semantic distance (CNN8; predicted from VC; shaded areas, 95% CI across five subjects). (**c**) Correlation coefficients between the rank and the semantic distance for all feature types/layers (predicted from VC). Correlation coefficients were calculated for each candidate set and target category, and then averaged across 1,000 repetitions of candidate selection and 50 target categories (error bars, 95% CI across five subjects; asterisks, one-sided *t* test after Fisher's Z transform, uncorrected $P < 0.05$).

For quantitative evaluation, we assessed the identification accuracy of seen and imagined objects when the number of candidate sets was two (Fig. 10a and b; see Supplementary Fig. 10 for identification accuracy as a function of the number of average samples). This analysis was performed with feature vectors for individual feature types/layers, and with concatenated feature vectors for CNN1–8 (8,000 units), HMAX1–3 (3,000 units), and all of the 13 feature types/layers (13,024 units).

The analysis revealed that both seen and imagined objects were successfully identified at a statistically significant level for most of the feature types/layers (one-sided $t$ test, uncorrected $P < 0.05$ except for CNN1, HMAX2, and HMAX3 under the imagery condition) with highest accuracy around mid-level features (CNN5–6). Furthermore, when the same analysis was performed for each ROI, above-chance accuracy was achieved for most of the feature–ROI combinations (Supplementary Fig. 11). Intriguingly, mid-level features decoded from higher ROIs were the most useful in identifying both seen and imagined object categories.

Additional analyses revealed that the pairs with larger semantic distances (*WordNet*'s path length) tended to have higher identification accuracies (Supplementary Fig. 14), consistent to the relations between the identified rank and semantic distance (Fig. 9c). Because we used the pre-trained CNN model, the 1,000 categories used in this model accidentally included 20 of the test categories in our study. However, the identification accuracy with the other 30 non-overlapping categories alone was qualitatively similar to the main results (Supplementary Fig. 15)

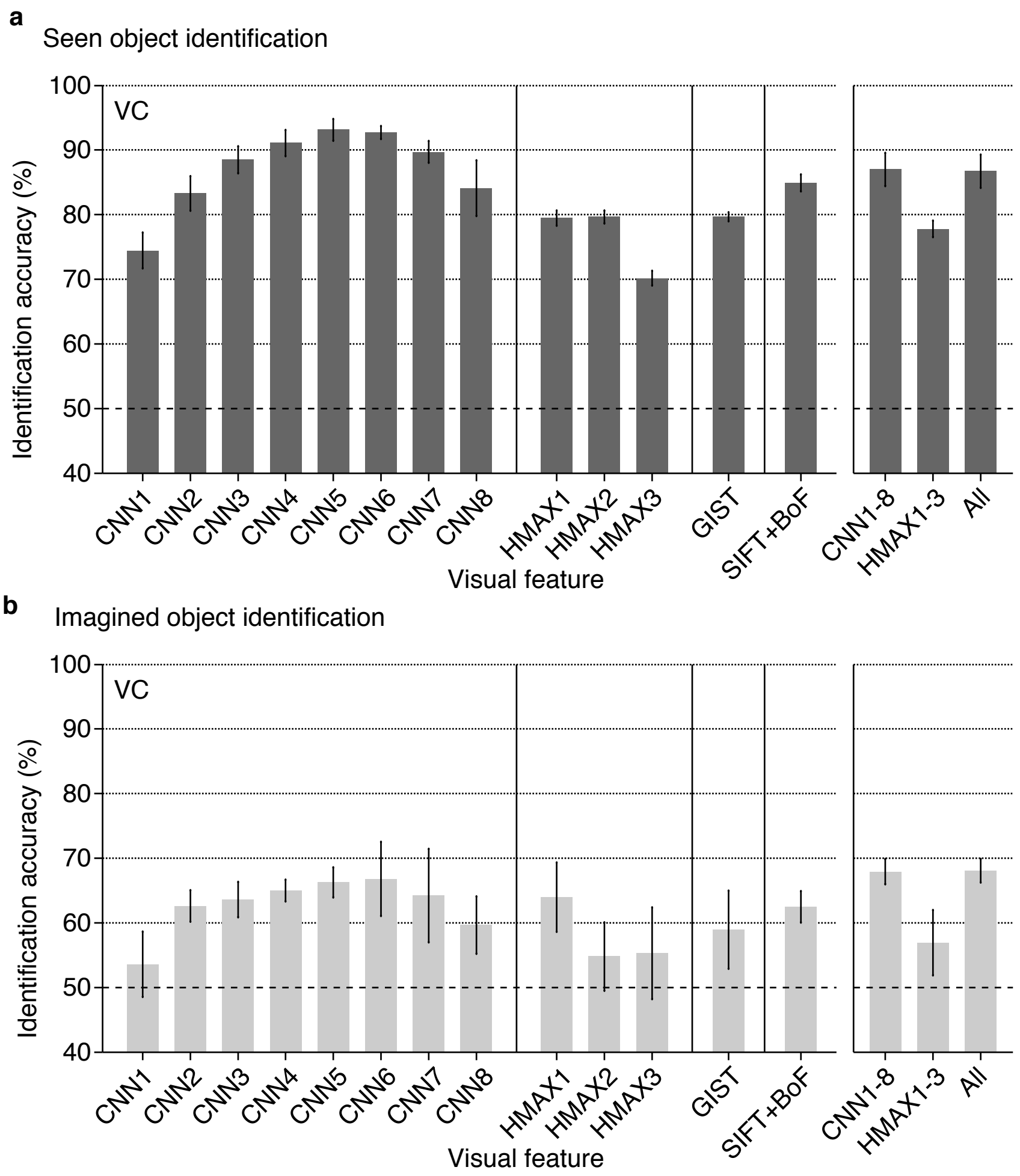

**Figure 10 | Identification accuracy.** Identification was performed for all combinations of one of the 50 test object categories and one of the 15,322 candidate categories (identification from two categories; predicted from VC: error bars, 95% CI across five subjects; dashed line, chance level, 50%). (**a**) Seen object identification accuracy. (**b**) Imagined object identification accuracy. While here the identification analyses were performed with the image feature decoders trained to predict image features of presented images, the category-average feature decoders trained with category-average features for training stimulus images can also be used. See Supplementary Fig. 12 and 13 for the results by the category-average feature decoders.

# Discussion

We have shown that via hierarchical visual feature representation, arbitrary object categories seen and imagined by subjects can be predicted from fMRI signals of the human visual cortex. The trained decoders successfully predicted the feature values of the presented images. Higher/lower-order visual features tended to be better predicted from fMRI signals in higher/lower cortical areas, respectively. Further, the decoders trained to predict feature vectors of presented images can be used to predict those of both seen and imagined object categories, enabling the identification of seen and imagined object categories despite not using the same categories for decoder training. Interestingly, mid-level features were the most useful in identifying object categories, suggesting the significant contributions of mid-level features to construct discriminative representations for object categories. Our results demonstrate that the decoding model trained on a limited set of object categories generalizes to decode arbitrary object categories, providing a proof of concept for generic object decoding. Moreover, successful predictions of category-average features at mid-to-high level CNN layers and object category identification from imagery-induced brain activity in mid-to-high level ROIs would suggest that mental imagery may recruit neural representations of visual features with intermediate complexity, which are elicited in visual perception, via progressive top-down processing.

Our analyses demonstrated that visual features extracted by computational models were successfully predicted from brain activity patterns (Fig. 3). The analysis revealed a hierarchical correspondence between the cortical hierarchy and the levels of visual feature representations. These results were consistent with the previous studies showing a high representational similarity between the top layer of a convolutional neural network and visual cortical activity in the inferior temporal (IT) cortex of humans[27,28] and non-human primates[25-27]. Moreover, several studies reported that the features from the middle layer of a hierarchical neural network can accurately predict V4 brain activity[25,26]. In addition, an explicit gradient for the feature complexity in the visual cortical hierarchy using the encoding approach has also been reported[28]. Our results were nearly equivalent to these findings, showing a homology between the hierarchies of individual subareas from lower to higher visual cortex and the CNN layers using the decoding approach. We thus confirmed the results supporting the theory that the CNN can be a good proxy for the hierarchical feed-forward visual system for object recognition.

In our analyses, we selected 1,000 feature units to reduce the computational cost, as some layers of the CNN (CNN1–7) originally had more than 100,000 units. It may have been possible that the selection of the 1,000 units biased the results. To verify that this was not the case, we repeated the same identification analysis by resampling units from the original feature populations and changing the number of units from 10 to 1,000 repeatedly (Supplementary Fig.16). The analysis demonstrated that the identification accuracies of most feature types/layers were almost saturated when several hundreds of units were used, and qualitatively similar results to our main results were obtained for different numbers of units.

Our results may be relevant to the long-standing debate as to whether mental imagery is symbolic (language-like) or depictive (picture-like)[39,40]. In our analysis, the decoders trained on brain activity induced by visual stimuli can generalize to predict the category-average feature vector of not only seen but also imagined object categories (Fig. 5), enabling the identification of imagined object categories using the feature vector decoded from brain activity during imagery (Fig. 10b). Previous studies have shown that the common neural representations are used during perception and mental imagery for low-level image properties[3-5,17], including information about orientation, spatial frequency, and retinotopic location, and also for high-level semantic and conceptual representations[8-10,12], depending on the tasks performed by subjects. Meanwhile, our analyses revealed progressive recruitments of multiple levels of hierarchical visual features, including features with intermediate complexity that go beyond low-level image properties and bridge the gap between pictorial and conceptual representations (Fig. 7). Our analyses quantify the top-down effects of hierarchical visual cortical activities during mental imagery, suggesting that feature-level representations elicited in visual perception were recruited during mental object imagery in a graded manner. These results reveal the nature of mental imagery as a type of top-down perception.

We have extended the modular decoding approach previously proposed in our visual image reconstruction study[16] to high-level object vision using the visual features with multiple levels of complexity (Fig. 4a). Since images depicting the same kinds of objects do not necessarily have pixel-wise similarity, complex and invariant features appear to be suitable for object identification. However, a simulation study suggests that intermediate-level rather than overly complicated visual features are better for object discrimination[41]. Our additional analysis with true feature vectors calculated from stimulus images (with no prediction errors; equivalent to generic object recognition in machine vision) also showed a slightly poorer identification with CNN8 than with CNN7 (Supplementary Fig. 17). Consistent with these observations, our results from decoded feature vectors showed the highest identification accuracy with mid-level rather than top-level CNN features. The results suggest the suitability of mid-level, or generic, features for discriminative object representation, possibly consistent with electrophysiological studies of the monkey inferior temporal cortex[37,42].

Our approach is relevant to a study that focused on semantic feature representations to establish relations between meanings of words and brain activities[43]. The study demonstrated decoding of arbitrary nouns thought by subjects using encoding models and statistics of word co-occurrence. Our work differs from that attempt in that we employed computational models that produced visual feature representations from images to establish the relations between brain activities in visual cortex and object categories, making it possible to address how hierarchical visual feature representations associated with visual objects are recruited during mental imagery. Furthermore, in the experiment of the previous study[43], they presented both line drawings and noun labels of concrete objects, and asked the subjects to perform the task of thinking about the properties of the target object. In contrast, in our image presentation experiment only visual images were presented without a cognitive task except for the one-back repetition task to keep subjects' attention, and in the imagery experiment, subjects were only required to imagine visual images of the target category without any additional tasks. Our experiments demonstrated that imagining about object images is sufficient to achieve generic decoding of imagined object categories utilizing the commonality of feature-level representations between perception and imagery.

In the present study, we used the decoding approach instead of the representational similarity analysis (RSA)[27,44-47] or encoding approaches[14,15,17,25,26,28] to link brain activities and visual features. While the RSA and encoding approaches can evaluate mass characteristics of each visual features (e.g., a specific layer of the CNN) associated with brain activity, our decoding approach can characterize individual feature unit in terms of the decodability from distributed brain activity patterns. As previously demonstrated, computational models with a high representational similarity to brain activity patterns in the inferior temporal cortex showed better categorization accuracy[27]. Thus, it may be possible to use the prediction accuracy of individual units as a guide to find effective units for better object recognition performance in machine vision. Indeed, our analysis revealed that the feature units better predicted from brain activity showed high category discriminability (Fig. 5). This finding is consistent with the conclusions of previous studies[25-27], and supports the notion that feature decoding accuracy may be a suitable guide for selecting better features for object recognition.

Our identification analyses revealed that highly ranked candidate categories tended to be semantically similar to the target categories for mid-to-high level CNN layers and SIFT+BoF (Fig. 9b and c). Therefore, even when the identification was incorrect, we were able to predict semantically similar categories to the target category. However, such a tendency was not observed with HMAX and GIST. Because GIST captures low-level image properties[24,29], it is not surprising that rank and the semantic distance were not correlated. On the other hand, HMAX was designed to model the higher-level visual system, but did not show positive correlations between rank and semantic distance. This may suggest that HMAX cannot capture the semantics of objects, which is consistent to the indications of several previous studies[27,44-47].

Our approach allows the decoding of arbitrary object categories not limited to those used for decoder training. This is a scientific and practical contribution, especially in situations where what kind of objects should be decoded is unknown. Because our approach can decode imagined categories, we may also be able to decode the contents of dreaming or daydreaming[13]. Reading contents of such spontaneously generated thinking would be beneficial in understanding the functions of such cognitive phenomena. Achieving this requires distinguishing the conceivable differences in neural representation between volitional and spontaneous mental imagery. This formulates a challenging problem in future work. In addition, our approach may provide a basis for a brain-based information retrieval system by translating brain activity into words or concepts. Using the outputs of decoders, our approach may create a query for an information retrieval system based on brain activity.

Furthermore the framework of directly predicting visual features from brain activity may be utilized for applications developed with deep neural networks. Recent advances of deep neural networks have enabled image reconstruction[48] and description generation[51] from feature patterns obtained by processing images through CNN. Combining these technologies with the brain decoding may extend previous reconstruction study[16] and the present work to produce richer outputs. Our results demonstrating the predictability of CNN features from the brain may then open the possibility to develop new technology for brain machine interface by combining the brain decoding and deep neural networks.

# Methods

### Subjects

Five healthy subjects (one female and four males, aged between 23 and 38) with normal or corrected-to-normal vision participated in the experiments. No statistical methods were used to determine the sample size but our sample size was chosen to match previous fMRI studies with a similar behavioral protocol. All subjects had considerable experience to participate fMRI experiments and were highly trained. All subjects provided written informed consent for their participation in the experiments, and the study protocol was approved by the Ethics Committee of ATR.

### Visual images

Images were collected from the online image database *ImageNet*[31] (2011, fall release), an image database where images are grouped according to the hierarchy in *WordNet*[38]. We selected two hundred representative object categories (synsets) for stimuli in the visual image presentation experiment. After excluding images with a width or height less than 100 pixels or aspect ratio over 1.5 or under 2/3, all remaining images in *ImageNet* were cropped to the center.

### Experimental design

We conducted two types of experiments: an image presentation experiment, and an imagery experiment. All visual stimuli were rear-projected onto a screen in an fMRI scanner bore using a luminance-calibrated LCD projector. Data from each subject were collected over multiple scanning sessions spanning approximately 2 months. In each experiment day, one consecutive session was conducted for two hours at the maximum. The subjects were given adequate time for rest between runs (every 3~10 minutes), and were allowed to take a break and stop the experiment on the day whenever they asked for.

The image presentation experiment consisted of two distinct types of sessions: training image sessions and test image sessions, each of which consisted of 24 and 35 separate runs (9 minutes 54 seconds for each run), respectively. Each run contained 55 stimulus blocks consisting of 50 blocks with different images and five randomly interspersed repetition blocks where the same image as in the previous block was presented. In each stimulus block an image (12 × 12 degree) was flashed at 2 Hz for 9 seconds. Images were presented on the center of the display with a central fixation spot. The color of the fixation spot changed from white to red for 0.5 seconds before each stimulus block began to indicate the onset of the block. Extra 33-second and 6-second rest periods were added to the beginning and end of each run, respectively. Subjects maintained steady fixation throughout each run, and performed a one-back repetition detection task on the images, responding with a button press for each repetition to maintain their attention on the presented images (mean task performance across five subjects; sensitivity = 0.930; specificity = 0.995). In the training image session, a total of 1,200 images from 150 object categories (eight images from each category) were each presented only once. In the test image session, a total of 50 images from 50 object categories (one image from each category) were presented 35 times each. Note that the categories in the test image session were not used in the training image session. The presentation order of the categories was randomized across runs.

In the imagery experiment, the subjects were required to visually imagine images from one of the 50 categories that were presented in the test image session of the image presentation experiment. Prior to the experiment, 50 image exemplars from each category were exposed to train the correspondence between object names and the visual images specified by the names. The imagery experiment consisted of 20 separate runs and each run contained 25 imagery blocks (10 minutes 39 seconds for each run). Each imagery block consisted of a 3-second cue period, a 15-second imagery period, a 3-second evaluation period, and a 3-second rest period. Extra 33-second and 6-second rest periods were added to the beginning and end of each run, respectively. During the rest periods, a white fixation spot was presented at the center of the display. The color of the fixation spot changed from white to red for 0.5 seconds to indicate the onset of the blocks from 0.8 seconds before each cue period began. During the cue period, words describing the names of the 50 categories presented in the test image session were visually presented around the center of the display (one target and 49 distractors). The position of each word was randomly changed across blocks to avoid contamination of cue-specific effects on the fMRI response during imagery periods. The word of the category to be imagined was presented with a red color (target) and the other words were presented in black (distractors). The onset and end of the imagery periods were signaled by beep sounds. The subjects were required to start imagining as many object images pertaining to the category described by the red word as possible. Their eyes were closed from the first beep to the second beep. After the second beep, the word of the target category was presented to allow the subjects evaluate the vividness of their mental imagery on a five-point scale (very vivid, fairly vivid, rather vivid, not vivid, cannot recognize the target) by a button press. The 25 categories in each run were pseudo-randomly selected from 50 categories such that the two consecutive runs contained all the 50 categories.

**Retinotopy experiment**

The retinotopy experiment followed the conventional protocol[50,51] using a rotating wedge and an expanding ring of a flickering checkerboard. The data were used to delineate the borders between each visual cortical area, and to identify the retinotopic map (V1–V4) on the flattened cortical surfaces of individual subjects.

**Localizer experiment**

We performed functional localizer experiments to identify the lateral occipital complex (LOC), fusiform face area (FFA), and parahippocampal place area (PPA) for each individual subject[52-54]. The localizer experiment consisted of four to eight runs and each run contained 16 stimulus blocks. In this experiment, intact or scrambled images (12 × 12 degree) of face, object, house, and scene categories were presented at the center of the screen. Each of eight stimulus types (four categories × two conditions) was presented twice per run. Each stimulus block consisted of a 15-second intact or scrambled stimulus presentation. The intact and scrambled stimulus blocks were presented successively (the order of the intact and scrambled stimulus blocks was random), followed by a 15-second rest period consisting of a uniform gray background. Extra 33-second and 6-second rest periods were added to the beginning and end of each run, respectively. In each stimulus block, 20 different images of the same type were presented for 0.3 seconds, followed by intervening 0.4-second-long blanks.

## MRI acquisition

fMRI data were collected using 3.0-Tesla Siemens MAGNETOM Trio a Tim scanner located at the ATR Brain Activity Imaging Center. An interleaved T2*-weighted gradient-EPI scan was performed to acquire functional images to covering the entire brain (image presentation, imagery, and localizer experiments: TR, 3,000 ms; TE, 30 ms; flip angle, 80 deg; FOV, 192 × 192 mm; voxel size, 3 × 3 × 3 mm; slice gap, 0 mm; number of slices, 50) or the entire occipital lobe (retinotopy experiment: TR, 2,000 ms; TE, 30 ms; flip angle, 80 deg; FOV, 192 × 192 mm; voxel size, 3 × 3 × 3 mm; slice gap, 0 mm; number of slices, 30). T2-weighted turbo spin echo images were scanned to acquire high-resolution anatomical images of the same slices used for the EPI (image presentation, imagery, and localizer experiments: TR, 7,020 ms; TE, 69 ms; flip angle, 160 deg; FOV, 192 × 192 mm; voxel size, 0.75 × 0.75 × 3.0 mm; retinotopy experiment: TR, 6,000 ms; TE, 57 ms; flip angle, 160 deg; FOV, 192 × 192 mm; voxel size, 0.75 × 0.75 × 3.0 mm). T1-weighted magnetization-prepared rapid acquisition gradient-echo (MP-RAGE) fine-structural images of the entire head were also acquired (TR, 2,250 ms; TE, 3.06 ms; TI, 900 ms; flip angle, 9 deg, FOV, 256 × 256 mm; voxel size, 1.0 × 1.0 × 1.0 mm).

## MRI data preprocessing

The first 9-second scans for experiments with TR = 3 seconds (image presentation, imagery, and localizer experiments) and 8-second scans for experiments with TR = 2 seconds (retinotopy experiment) of each run were discarded to avoid MRI scanner instability. The acquired fMRI data underwent three-dimensional motion correction by SPM5 (http://www.fil.ion.ucl.ac.uk/spm). The data were then coregistered to the within-session high-resolution anatomical image of the same slices used for EPI and subsequently to the whole-head high-resolution anatomical image. The coregistered data were then reinterpolated by 3 × 3 × 3 mm voxels.

For the data from the image presentation experiment and imagery experiment, after within-run linear trend removal, voxel amplitudes were normalized relative to the mean amplitude of the entire time course within each run. The normalized voxel amplitudes from each experiment were then averaged within each 9-second stimulus block (three volumes; image presentation experiment) or within each 15-second imagery period (five volumes; imagery experiment) respectively (unless otherwise stated) after shifting the data by 3 seconds (one volume) to compensate for hemodynamic delays.

**Region of interest (ROI) selection**

V1, V2, V3, and V4 were delineated by the standard retinotopy experiment[50,51]. The retinotopy experiment data were transformed to Talairach coordinates and the visual cortical borders were delineated on the flattened cortical surfaces using BrainVoyager QX (http://www.brainvoyager.com). The voxel coordinates around the gray-white matter boundary in V1–V4 were identified and transformed back into the original coordinates of the EPI images. The voxels from V1–V3 were combined as the *lower visual cortex* (LVC). The LOC, FFA, and PPA were identified using conventional functional localizers[52-54]. The localizer experiment data were analyzed using SPM5. The voxels showing significantly higher responses to objects, faces, or scenes than for scrambled images (two-sided *t* test, uncorrected $P < 0.05$ or 0.01) were identified, and defined as LOC, FFA, and PPA, respectively. A contiguous region covering LOC, FFA, and PPA was manually delineated on the flattened cortical surfaces, and the region was defined as the *higher visual cortex* (HVC). Voxels overlapping with LVC were excluded from HVC. Voxels from V1–V4 and HVC were combined to define the *visual cortex* (VC). In the regression analysis, voxels showing the highest correlation coefficient with the target variable in the training image session were selected to predict each feature (at most 500 voxels for V1, V2, V3, V4, LOC, FFA, and PPA; 1,000 voxels for LVC, HVC, and VC).

## Visual features

We used four types of computational models: a convolutional neural network (CNN)[20], HMAX[21-23], GIST[24], and scale invariant feature transform (SIFT)[18] combined with "Bag of Features (BoF)"[16] to construct visual features from images. The features with a model-training phase (HMAX and SIFT+BoF) used 1,000 images belonging to the categories used in the training image session (150 categories) for training. Each model is further described in the following subsections.

### *Convolutional neural network (CNN)*

We used the *MatConvNet* implementation (http://www.vlfeat.org/matconvnet/) of the CNN model[20], which was trained with images in *ImageNet*[31] to classify 1,000 object categories. The CNN consisted of five convolutional layers and three fully-connected layers. We randomly selected 1,000 units in each of the first to seventh layers and used all 1,000 units in the eighth layer.

We represented each image by a vector of those units' outputs, and named them as CNN1–CNN8, respectively.

### *HMAX*

HMAX[21-23] is a hierarchical model that extends the simple and complex cells of Hubel and Wiesel[55,56], and computed features through hierarchical layers. These layers consist of an image layer and six subsequent layers (S1, C1, S2, C2, S3, and C3), which are built from the previous ones by alternating template matching and max operations. In the calculations at each layer, we employed the same parameters as in a previous study[22], except that the number of features in layer C2 and C3 was set to 1,000. We represented each image by a vector of the three types of HMAX features, which consisted of 1,000 randomly selected outputs of units in layers S1, S2 and C2, and all 1,000 outputs in layer C3. We defined these outputs as HMAX1, HMAX2, and HMAX3, respectively.

### *GIST*

GIST is a model developed for the computer-aided scene categorization task[24]. To compute GIST, an image was first converted to gray-scale and resized to have a maximum width of 256 pixels. Next, the image was filtered using a set of Gabor filters (16 orientations, 4 scales). After that, the filtered images were segmented by a four by four grid (16 blocks), and then the filtered outputs within each block were averaged to extract 16 responses for each filter. The responses from multiple filters were concatenated to create a 1,024 dimensional feature vector for each image (16 [orientations] × 4 [scales] × 16 [blocks] = 1,024).

### *SIFT with BoF (SIFT+BoF)*

The visual features using the SIFT with BoF approach were calculated from SIFT descriptors. We computed SIFT descriptors from the images using the VLFeat[57] implementation of dense SIFT. In the BoF approach, each component of the feature vector (visualwords) is created by vector-quantizing extracted descriptors. Using about one million SIFT descriptors calculated from an independent training image set, we performed k-means clustering to create a set of 1,000 visualwords. The SIFT descriptors extracted from each image were quantized into visualwords using the nearest cluster center, and the frequency of each visualword was calculated to create a BoF histogram for each image. Finally, all of the histograms obtained through the above processing underwent L-1 normalization to become unit norm vectors. Consequently, features from SIFT with BoF approach are invariant to image scaling, translation, and rotation, and are partially invariant to illumination changes and affine or 3D projection.

## Visual feature decoding

We constructed decoding models to predict the visual feature vectors of seen objects from fMRI activity using a linear regression function. Here, we used sparse linear regression (SLR; http://www.cns.atr.jp/cbi/sparse_estimation/index.html)[32] that can automatically select the important features for prediction. The sparse estimation is known to perform well when the dimensionality of the explanatory variable is high, as is the case with fMRI data[58].

Given an fMRI sample $\mathbf{x} = \{x, \dots, x_d\}^{\mathrm{T}}$ consisting of the activities of $d$ voxels' as input, the regression function can be expressed by

$$y(\mathbf{x}) = \sum_{i=1}^{d} w_i x_i + w_0,$$

where $x_i$ is a scalar value specifying the fMRI amplitude of the voxel $i$, $w_i$ is the weight of voxel $i$, and $w_0$ is the bias. For simplicity, the bias $w_0$ is absorbed into the weight vector such that $\mathbf{w} = \{w_0, \dots, w_d\}^{\mathrm{T}}$. The dummy variable $x_0 = 1$ is introduced into the data such that $\mathbf{x} = \{x_0, \dots, x_d\}^{\mathrm{T}}$. Using this function, we modeled the *lth* component of each visual feature vector as a target variable $t_l$ ($l \in \{1, \dots, L\}$) that is explained by the regression function $y(\mathbf{x})$ with additive Gaussian noise as described by

$$t_l = y(\mathbf{x}) + \epsilon$$

where $\epsilon$ is a zero mean Gaussian random variable with noise precision $\beta$.

Given a training data set, SLR computes the weights for the regression function such that the regression function optimizes an objective function. To construct the objective function, we first express the likelihood function by

$$P(\mathbf{t}_l | \mathbf{X}, \mathbf{w}, \beta) = \prod_{n=1}^{N} \frac{1}{(2\pi)^{1/2}} \beta^{1/2} exp\left\{-\frac{1}{2}\beta(t_{ln} - \mathbf{w}^{\mathrm{T}}\mathbf{x}_n)^2\right\},$$

where $N$ is the number of samples, $\mathbf{X}$ is an $N \times (d+1)$ fMRI data matrix whose *n*th row is the $d+1$-dimensional vector $\mathbf{x}_n$, and $\mathbf{t}_l = \{t_{l1}, \dots, t_{ln}\}^{\mathrm{T}}$ are the samples of a component of the visual feature vector.

We performed Bayesian parameter estimation, and adopted the automatic relevance determination (ARD) prior[32] to introduce sparsity into the weight estimation. We considered the estimation of the weight parameter $\mathbf{w}$ given the training data sets $\{\mathbf{X}, \mathbf{t}_l\}$. We assumed a Gaussian distribution prior for the weights $\mathbf{w}$ and non-informative priors for the weight precision parameters $\boldsymbol{\alpha} = \{\alpha_0, \ldots \alpha_d\}^{\mathbf{T}}$ and the noise precision parameter $\beta$, which are described as

$$P_0(\mathbf{w}|\boldsymbol{\alpha}) = \prod_{i=0}^{d} \frac{1}{(2\pi)^{1/2}} \alpha_i^{1/2} exp\left\{-\frac{1}{2} \alpha_i \mathrm{w}_i^2\right\},$$

$$P_0(\boldsymbol{\alpha}) = \prod_{i=0}^{d} \frac{1}{\alpha_i},$$

$$P_0\left(\frac{1}{\beta}\right) = \frac{1}{\beta}.$$

In the Bayesian framework, we consider the joint probability distribution of all the estimated parameters, and the weights can be estimated by evaluating the following joint posterior probability of $\mathbf{w}$:

$$P(\mathbf{w}, \boldsymbol{\alpha}, \beta|\mathbf{X}, \mathbf{t}_l) = \frac{P(\mathbf{t}_l, \mathbf{w}, \boldsymbol{\alpha}, \beta|\mathbf{X})}{\int dwdad\beta\ P(\mathbf{t}_l, \mathbf{w}, \boldsymbol{\alpha}, \beta|\mathbf{X})} = \frac{P(\mathbf{t}_l|\mathbf{X}, \mathbf{w}, \beta)P_0(\mathbf{w}|\boldsymbol{\alpha})P_0(\boldsymbol{\alpha})P_0(\beta)}{\int dwdad\beta\ P(\mathbf{t}_l, \mathbf{w}, \boldsymbol{\alpha}, \beta|\mathbf{X})}.$$

Given that the evaluation of the joint posterior probability $P(\mathbf{w}, \boldsymbol{\alpha}, \beta|\mathbf{X}, \mathbf{t}_l)$ is analytically intractable, we approximate it using the variational Bayesian method[32,59,60]. The results obtained

using the standard linear regression model with maximum likelihood estimation were qualitatively similar to those obtained using our Bayesian sparse linear regression model.

We trained linear regression models that predict a feature vector of individual feature types/layers for seen object categories given fMRI samples in the training image session. For test datasets, fMRI samples corresponding to the same categories (35 samples in the test image session, 10 samples in the imagery experiment) were averaged across trials to increase the signal to noise ratio of the fMRI signals. Using the learned models, we predicted feature vectors of seen/imagined objects from averaged fMRI samples to construct one predicted feature vector for each of the 50 test categories.

**Synthesizing preferred images using activation maximization**

We used the activation maximization method to generate preferred images for individual units in each CNN layer[33–36]. Synthesizing preferred images starts from a random image and optimizes the image to maximally activate a target CNN unit by iteratively calculating how the image should be changed via backpropagation. This analysis was implemented using custom software written in MATLAB based on Python codes provided by the blog posts (Mordvintsev, A., Olah, C., Tyka, M., DeepDream - a code example for visualizing Neural Networks, https://github.com/google/deepdream, 2015; Øygard, A.M.,Visualizing GoogLeNet Classes, https://github.com/auduno/deepdraw, 2015).

## Identification analysis

In the identification analyses, seen/imagined object categories were identified using the visual feature vectors decoded from fMRI signals. Prior to the identification analysis, visual feature vectors were computed for all of the preprocessed images in all of the categories (15,372 categories in *ImageNet*[31]) except for those used in the fMRI experiments and their hypernym/hyponym categories and those used for visual feature model training (HMAX and SIFT+BoF). The visual feature vectors of individual images were averaged within each category to create category-average feature vectors for all of the categories to form the candidate set. We computed the Pearson's correlation coefficients between the decoded and the category-average feature vectors in the candidate sets. To quantify the accuracy, we created candidate sets consisting of the seen/imagined categories and the specified number of randomly selected categories. None of the categories in the candidate set were used for decoder training. Given a decoded feature vector, category identification was conducted by selecting the category with the highest correlation coefficient among the candidate sets.

## Statistics

In the main analysis, to test the statistical significance, we applied a $t$ test to test whether the mean of the correlation coefficients and the mean of the identification accuracies across subjects exceed the chance level (0 for correlation coefficient, and 50% for identification accuracy). For correlation coefficients, Fisher's Z transform were applied before the statistical tests. Before every $t$ test, we performed the Shapiro-Wilk test to check the normality, and we confirmed the null hypothesis that the data came from a normal distribution was not rejected for all cases ($P > 0.01$).

## Data and code availability

The experimental data and codes used in the present study are available from the corresponding author upon request.

## Acknowledgements

The authors thank Yoichi Miyawaki, Yasuhito Sawahata, Makoto Takemiya, and Kei Majima for helpful comments on the manuscript; and Mitsuaki Tsukamoto for help with data collection. Guohua Shen for help with implementing codes for generating preferred images. This research was supported by grants from JSPS KAKENHI Grant number JP26119536, JP26870935, JP15H05920, JP15H05710, a contract with the Ministry of Internal Affairs and Communications entitled "Novel and innovative R&D making use of brain structures", ImPACT Program of Council for Science, Technology and Innovation (Cabinet Office, Government of Japan), and the New Energy and Industrial Technology Development Organization (NEDO).

## Author contributions

TH and YK designed the study. TH performed experiments and analyses. TH and YK wrote the paper.

## Competing financial interests

The authors declare that they have no competing financial interests.

## Supplementary Information for

# Generic Decoding of Seen and Imagined Objects using Hierarchical Visual Features

Tomoyasu Horikawa and Yukiyasu Kamitani

### Table of contents

Supplementary figures

# Supplementary Figures

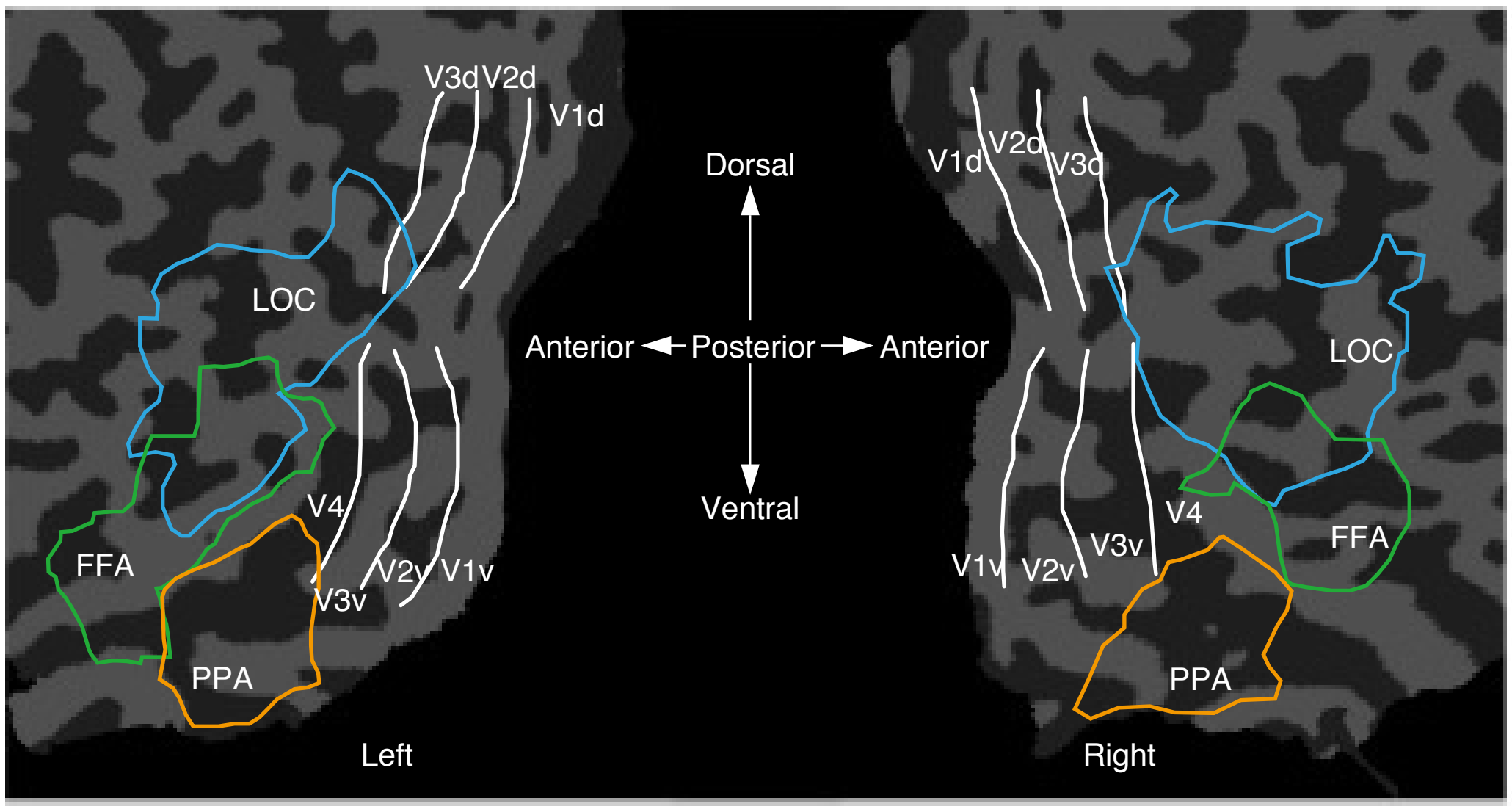

**Supplementary Figure 1 | Definitions of regions of interest on flattened cortex.** The individual ROIs of Subject 2 are shown on the flattened cortex. A contiguous region covering the LOC, FFA, and PPA was manually delineated on the flattened cortical surface, and the region was defined as the "higher visual cortex" (HVC). The voxels overlapping with the "lower visual cortex" (LVC, V1–V3) were excluded from ROI for the HVC. For individual ROIs voxels near the area border were included in both ROIs.

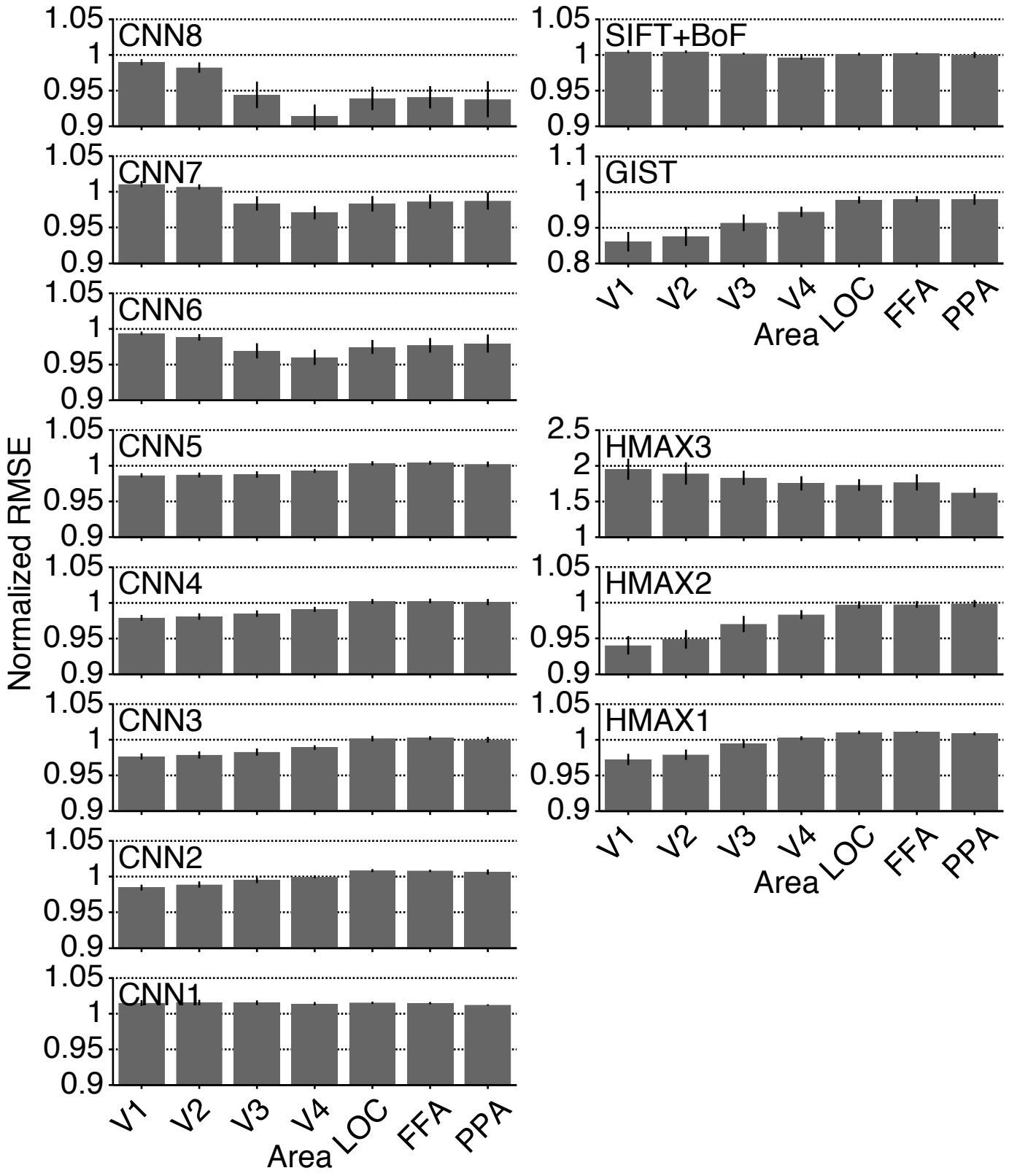

**Supplementary Figure 2 | Image feature decoding accuracy evaluated by normalized root mean square error.** For each feature unit, the root mean square error (RMSE) between true and predicted values were calculated over 50 test categories, and then normalized by the standard deviation of the true values. The normalized RMSE (nRMSE) was averaged for each combination of feature types/layers and ROIs (error bars, 95% CI across five subjects). The range of the horizontal axis was changed for each visual feature type/layer for display purposes. This analysis replicated a general trend observed in the results based on correlation coefficients (Fig. 3b), showing that the higher-order visual features tended to be better predicted from fMRI signals in higher rather than lower ROIs, and that lower-order visual features tended to be better predicted from fMRI signals in lower rather than higher ROIs (ANOVA, interaction between visual feature type/layer and ROI, $P < 0.01$). However, nRMSE showed a different pattern of accuracy from correlation when compared across feature types/layers. For example, HMAX3 showed the worst accuracy in nRMSE for all ROIs, while it attained a higher accuracy than several CNN features and SIFT+BoF in correlation.

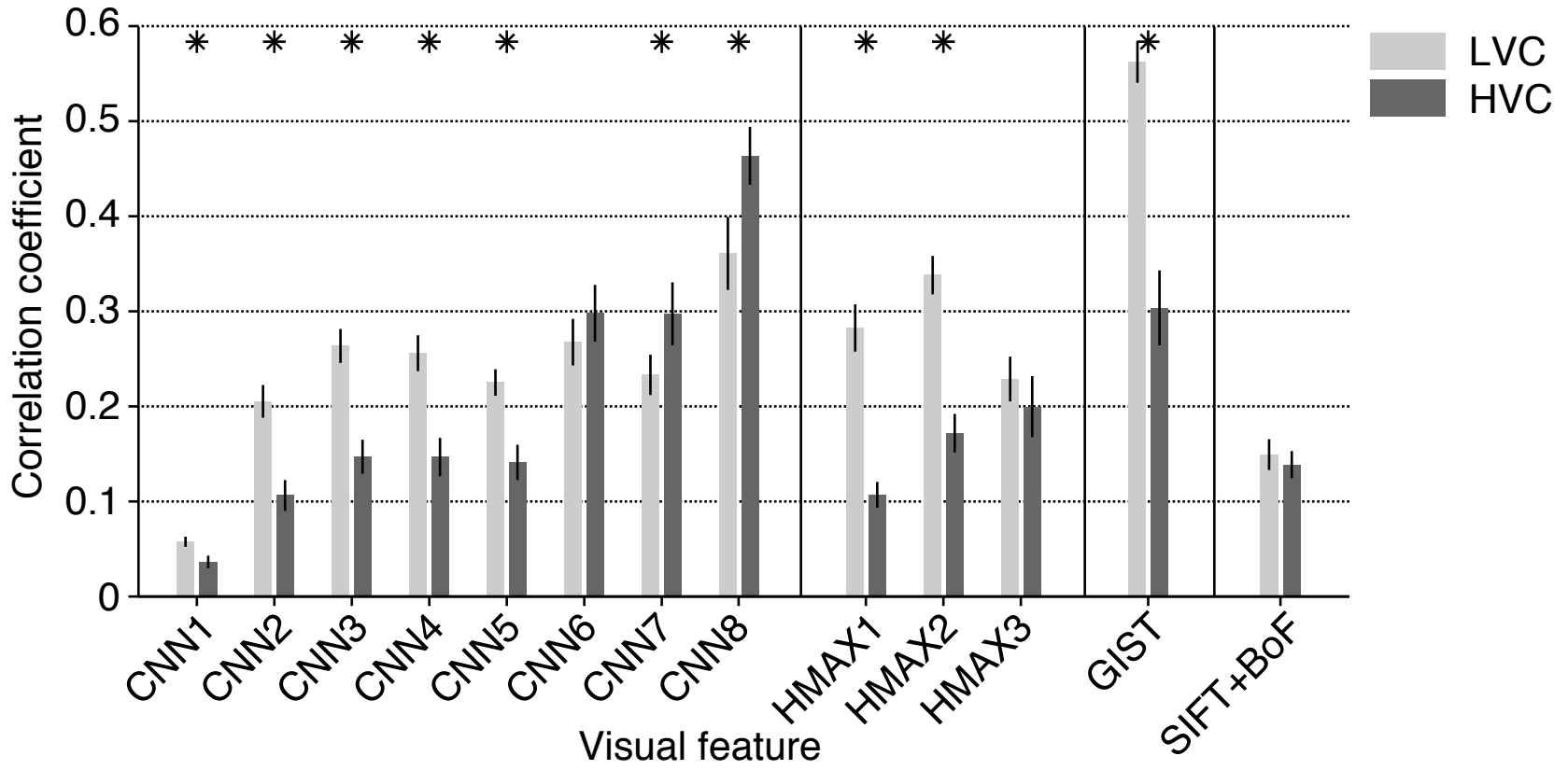

**Supplementary Figure 3 | Image feature decoding accuracy obtained by decoders trained with brain activities from lower and higher ROIs.** The image feature decoding accuracy obtained from decoders trained on brain activity patterns from the LVC and HVC are shown (error bars, 95% CI across five subjects). The analyses showed that the decoders trained on the LVC activity outperformed those trained on the HVC activity in CNN1–5, HMAX1 and 2, and GIST, while the opposite was observed in CNN7 and CNN8 (asterisk, two-sided *t* test, uncorrected $P < 0.01$; ANOVA, interaction between visual feature type/layer and ROI, $P < 0.01$, for both of the CNN feature set and HMAX feature set). The decoding accuracy between decoders of the LVC and HVC did not differ with a statistically significant level in CNN6, HMAX3, and SIFT+BoF (two-sided *t* test, uncorrected $P > 0.01$). These results characterized the visual feature types/layers with respect to the levels of visual cortical hierarchy. Before *t* test, we performed an *F* test to check the equality of variances between the results from the LVC and HVC, and we confirmed that the null hypothesis that the data for the LVC and HVC have the same variance was not rejected for all feature types/layers ($P > 0.05$).

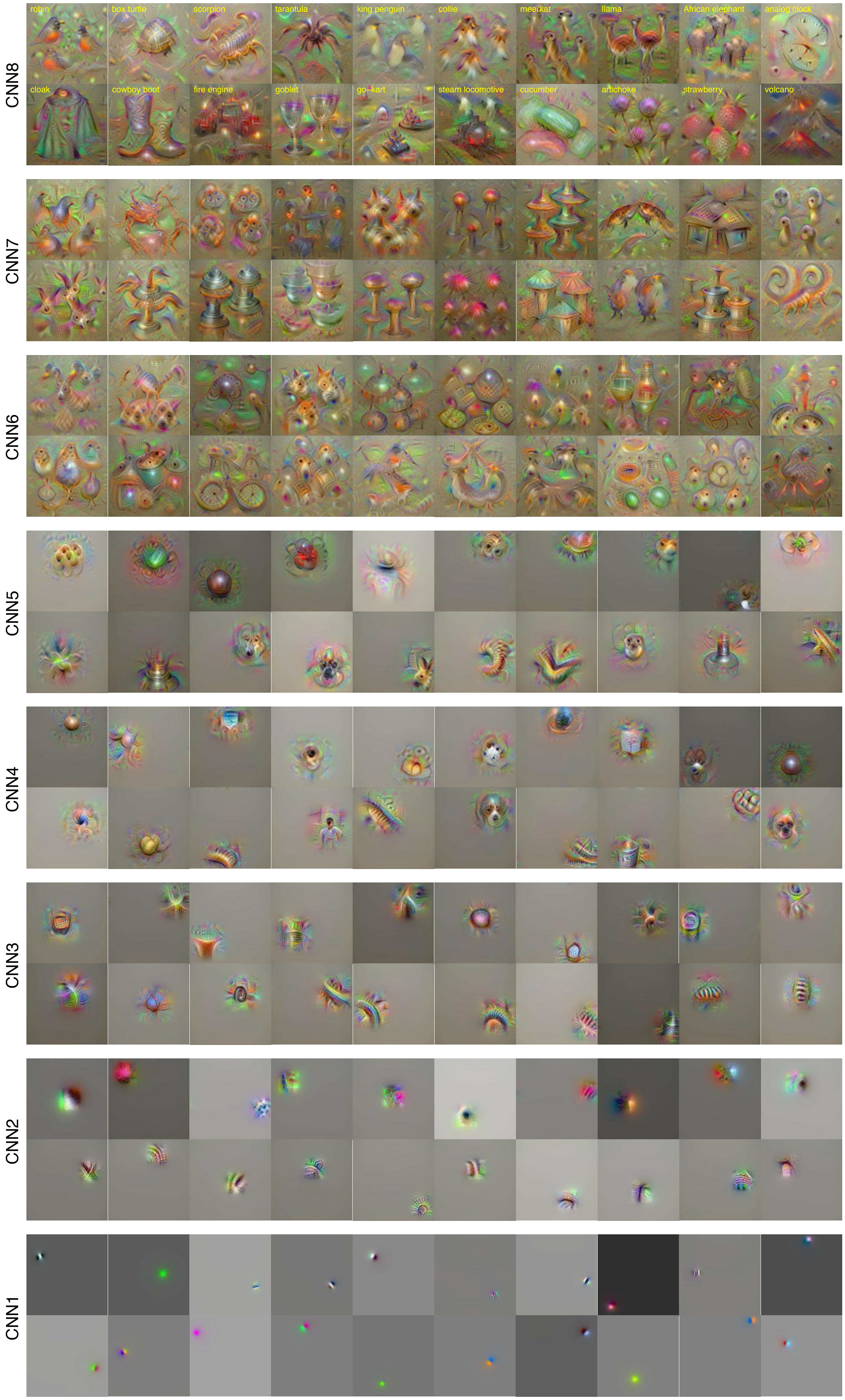

**Supplementary Figure 4 | Examples of preferred images for individual units in CNN layers.** Examples of preferred images synthesized for each of randomly selected twenty units in each CNN layer are shown. Category names of individual units in the CNN8 are shown on the left top of the images. Because the CNN6–8 are fully-connected layers, position information is lost for these layers.

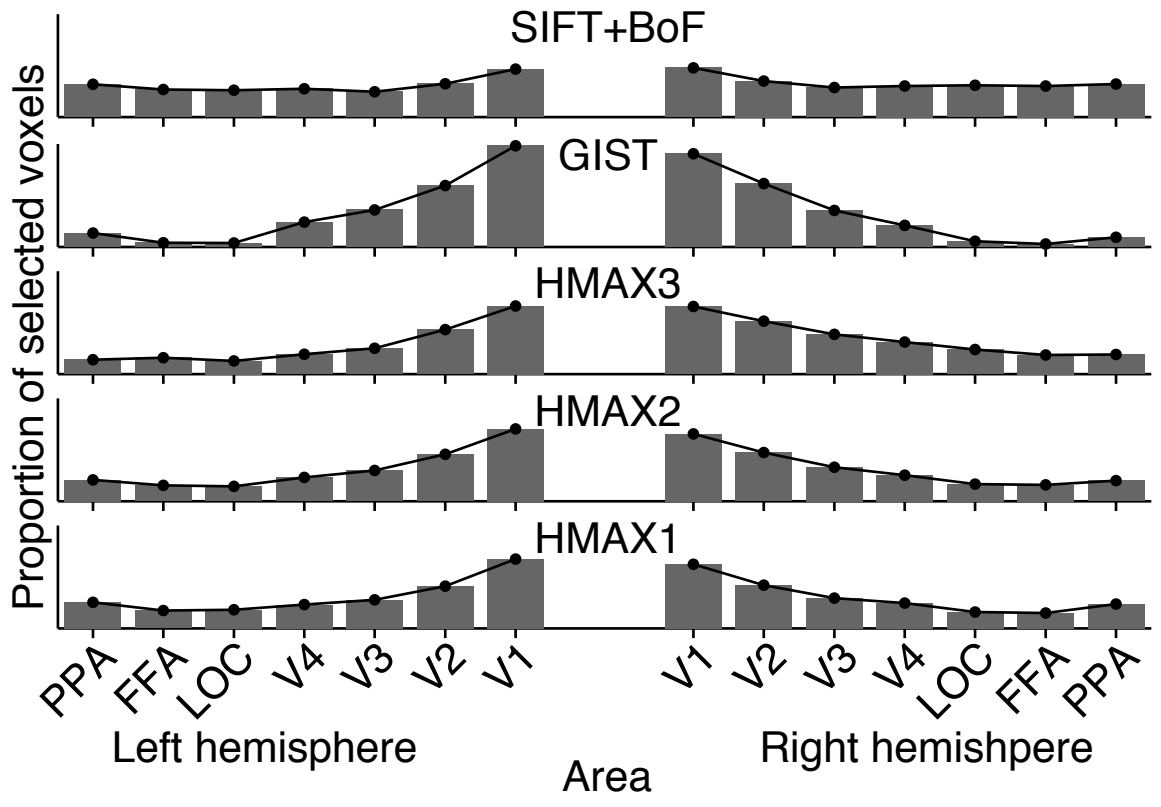

**Supplementary Figure 5 | Distributions of selected voxels across individual subareas for HMAX, GIST, and SIFT+BoF**. Distributions of selected voxels used for predictions of each visual feature type are shown for the HMAX, GIST, and SIFT+BoF (five subjects averaged, predicted from VC).

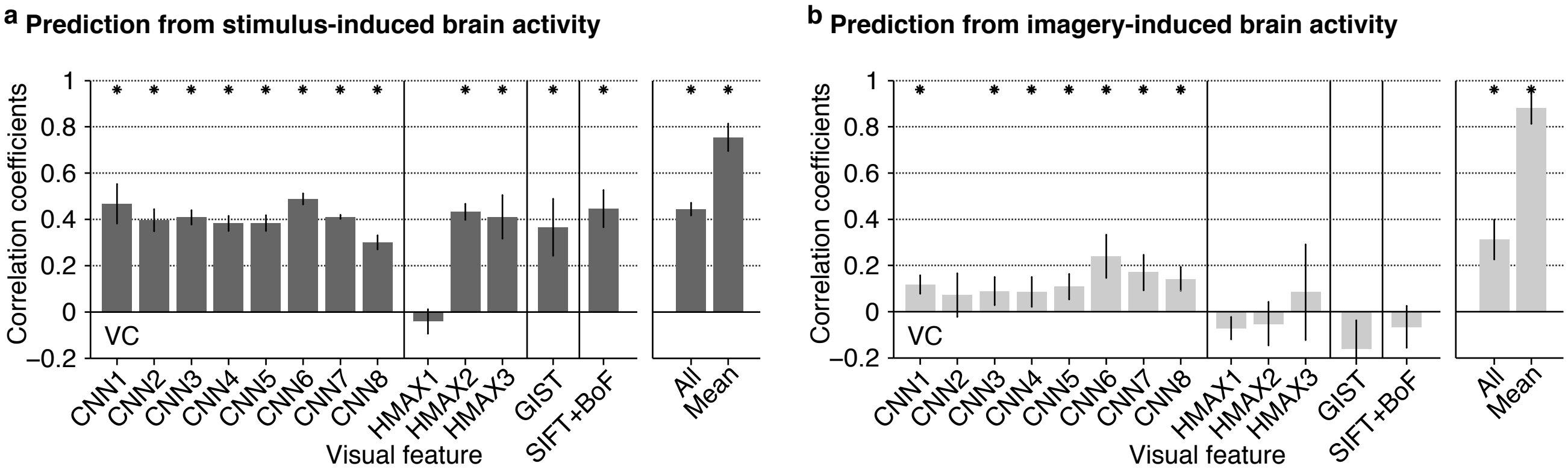

**Supplementary Figure 6 | Relation between category discriminability and prediction accuracy of category-average features.** The same analysis for Fig. 5c was performed with correlation coefficients between the values of the category-average features and the predicted features for seen and imagined conditions (predicted from VC by image feature decoders; cf., Fig. 6). Correlation coefficients between the category discriminability and the category-average feature decoding accuracy are shown for the seen and imagined conditions (error bars, 95% CI across five subjects; asterisks, one-sided *t* test after Fisher's Z transform, uncorrected $P < 0.05$). (**a**) Correlation coefficients obtained by predicting features from stimulus-induced brain activity. (**b**) Correlation coefficients obtained by predicting features from imagery-induced brain activity.

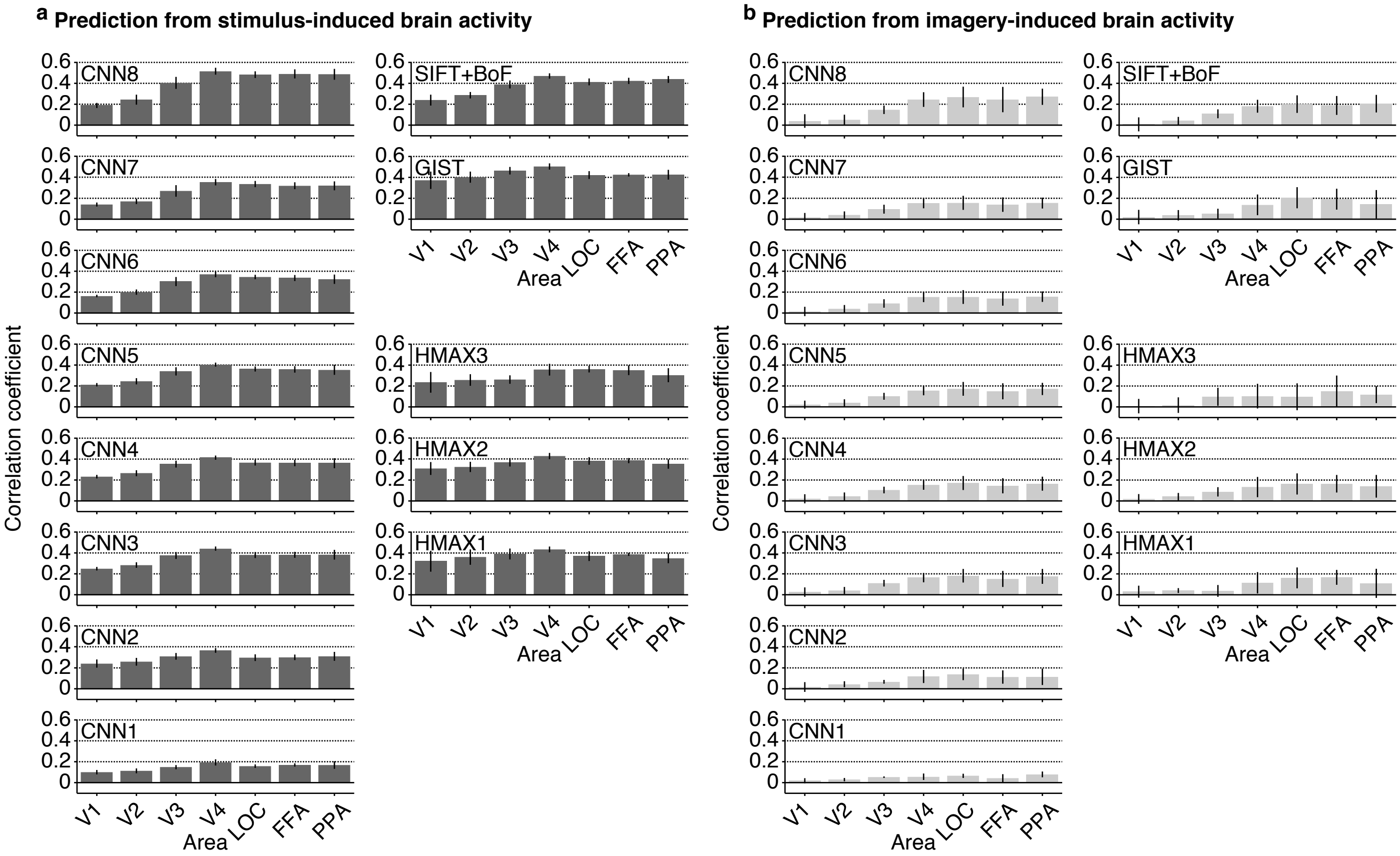

**Supplementary Figure 7 | Prediction of category-average features from stimulus- and imagery-induced brain activity by category-average feature decoders.** (**a**) Correlation coefficients with predicted features from stimulus-induced brain activity. (**b**) Correlation coefficients with predicted features from imagery-induced brain activity. Mean correlation coefficients are shown for each feature type/layer and ROI (error bars, 95% CI across five subjects).

**a Image feature decoders**

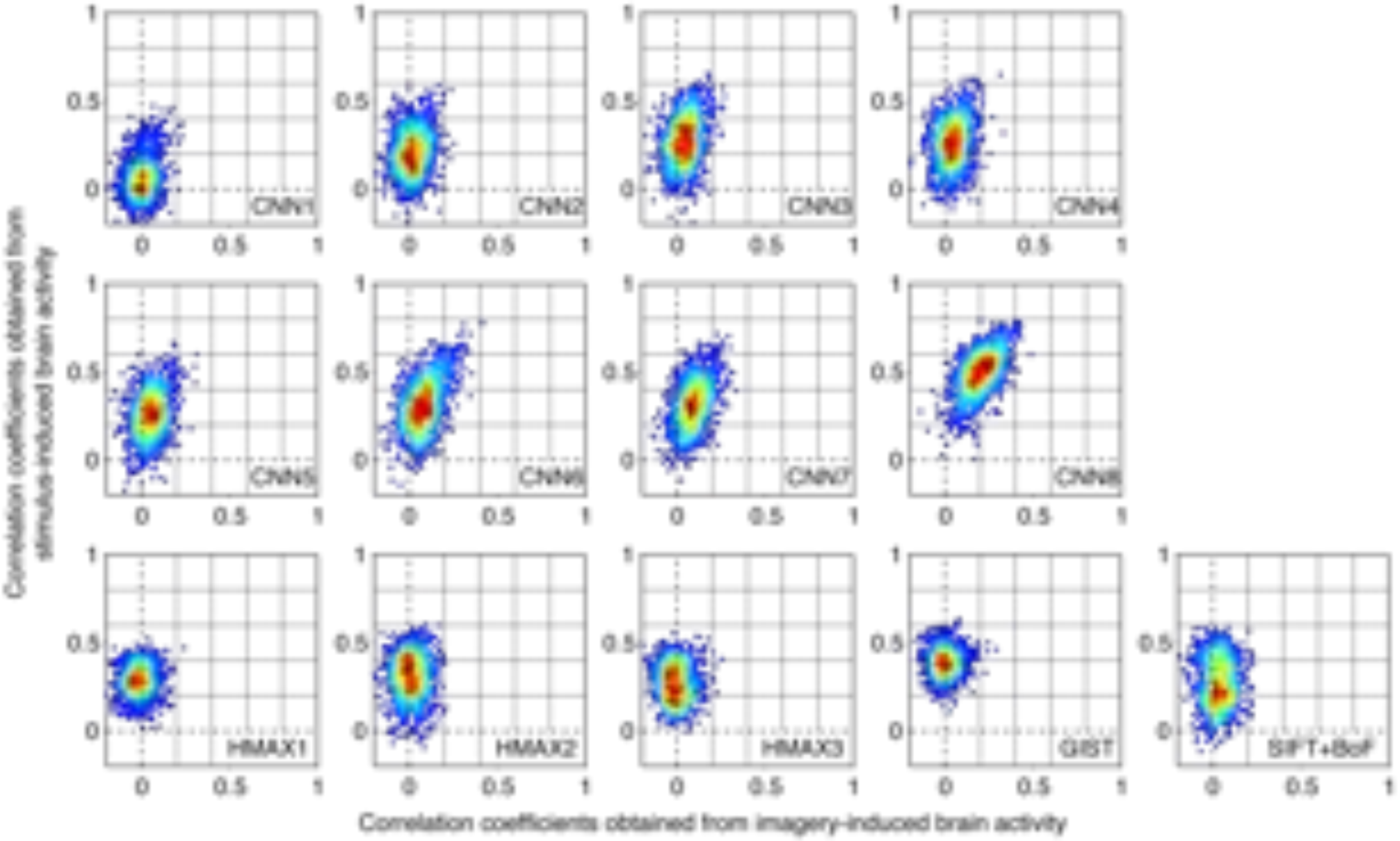

**b Category-average feature decoders**

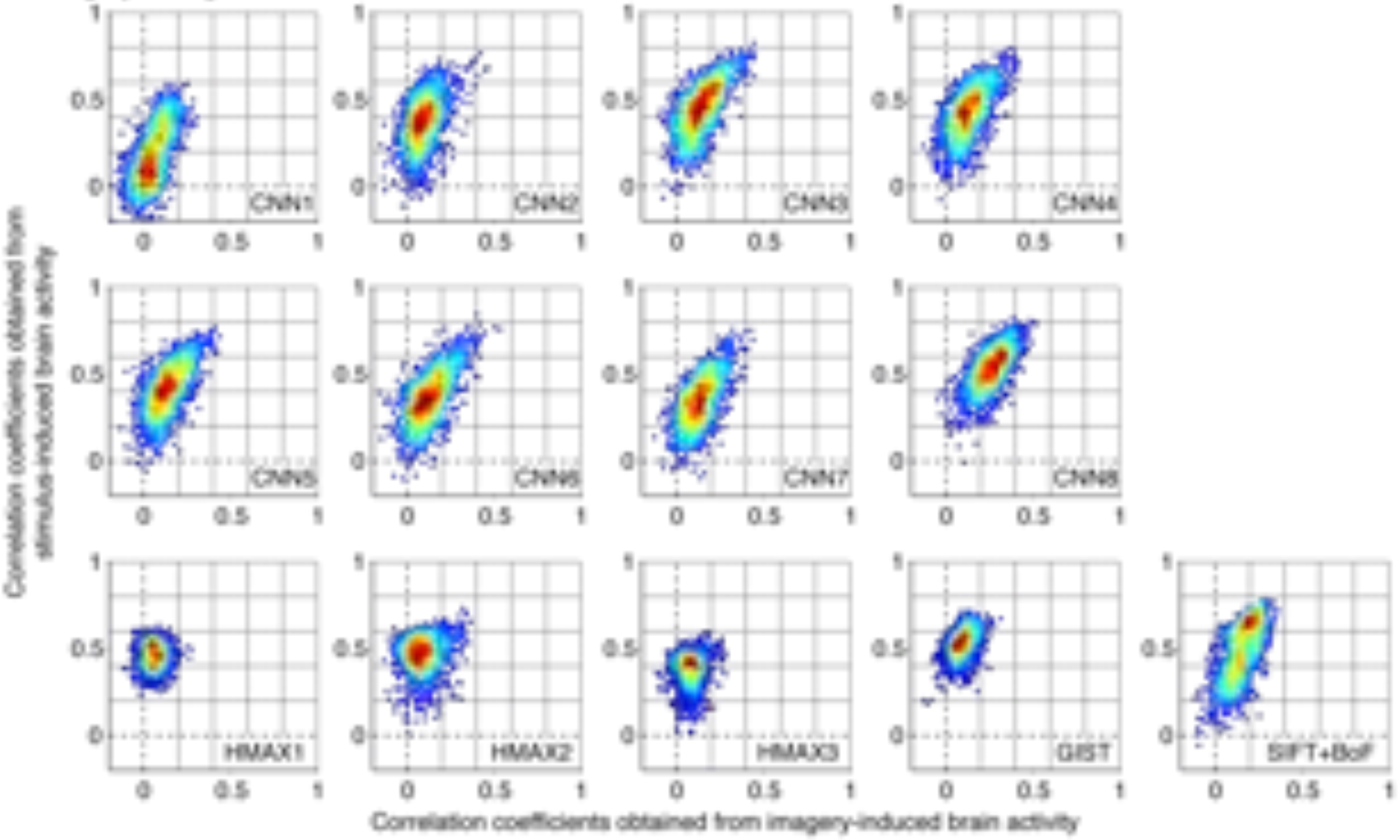

**Supplementary Figure 8 | Distributions of correlation coefficients between the predicted and the category-average feature values for seen and imagined conditions.** Scatterplots of correlation coefficients between the predicted and the category-average feature values for the seen (vertical axis) and imagined (horizontal axis) conditions are shown for ~1,000 feature units. (**a**) Distributions obtained by the image feature decoders. (**b**) Distributions obtained by the category-average feature decoders. Each dot denotes the averaged correlation coefficients across five subjects (predicted from VC) for each feature unit. The color indicates the density of the dots. Although the mean correlations spanned from around 0.1 to 0.5 for the seen condition (Fig. 6a and Supplementary Fig. 7a) and from around 0.0 to 0.2 for the imagined condition (Fig. 6b and Supplementary Fig. 7b), the correlations of individual units are rather broadly distributed. A subset of units with high correlations may substantially contribute to object category decoding.

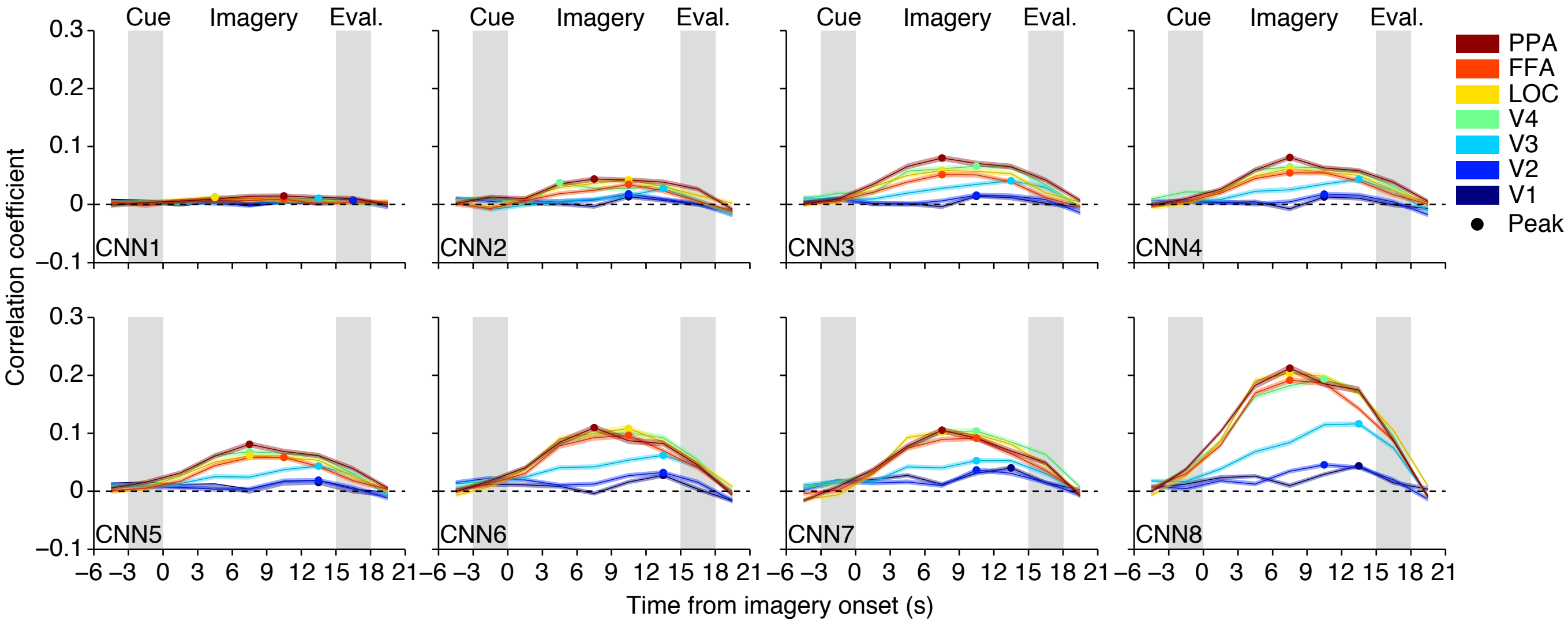

**Supplementary Figure 9 | Time course of feature prediction from imagery-induced brain activity for individual CNN layers.** At each time point/volume around the task period, correlation coefficients were calculated between the predicted and the category-average feature values for the series of test trials (five subject averaged; shaded areas, 95% CI across feature units; filled circles, peak timing). Prediction from imagery-induced brain activity in individual ROIs are shown for individual CNN layers.

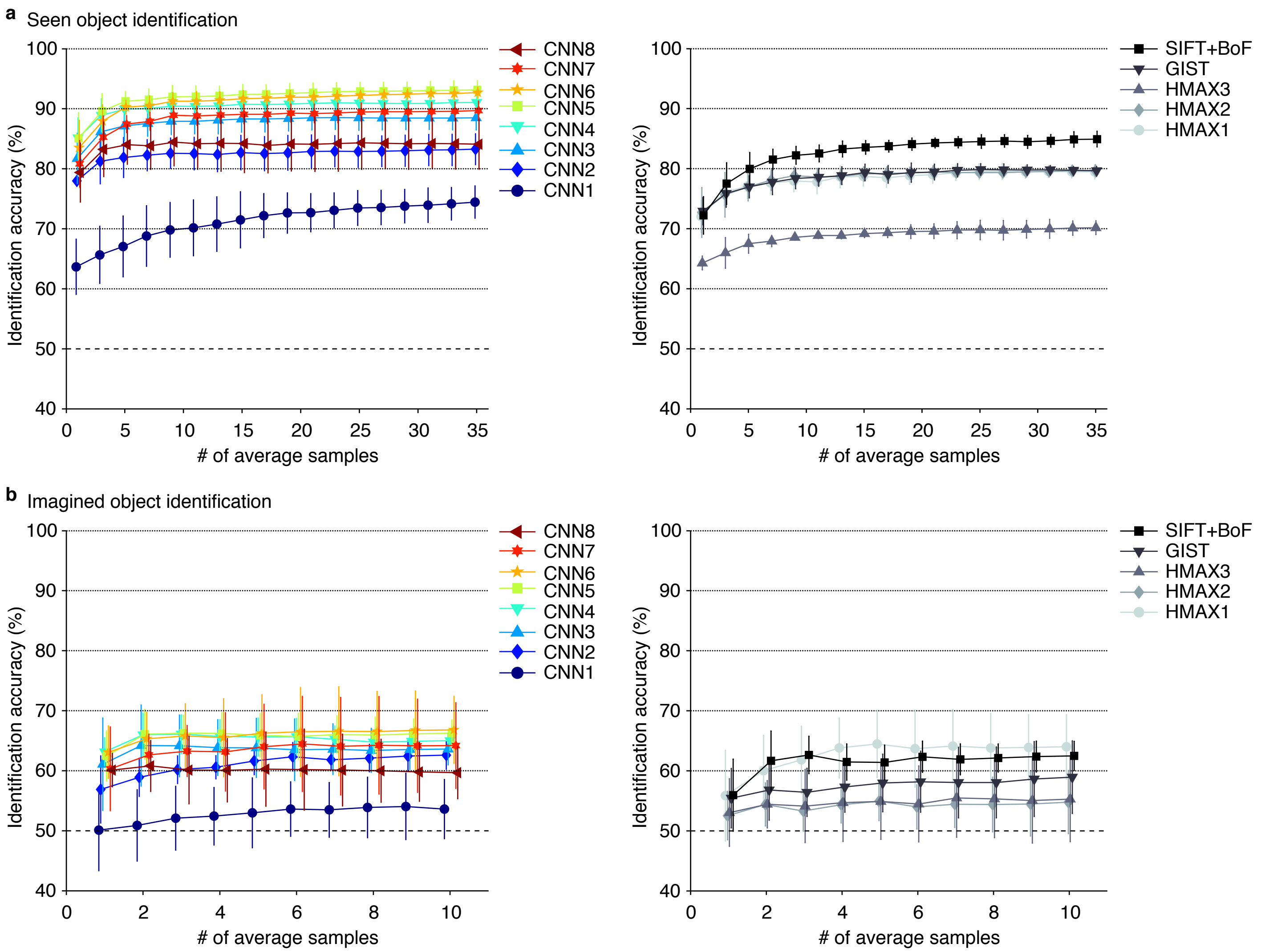

**Supplementary Figure 10 | Identification accuracy as a function of the number of average samples.** Identification accuracies as a function of the number of average samples are shown (identification from two categories; predicted from VC by image feature decoders; error bars, 95% CI across five subjects; dashed line, chance level, 50%). (**a**) Seen object identification accuracy. The identification accuracy gradually improves with the number of average samples but saturates at fewer than ten samples for most feature types/layers. (**b**) Imagined object identification accuracy. Nearly equivalent accuracies are observed even without averaging multiple samples.

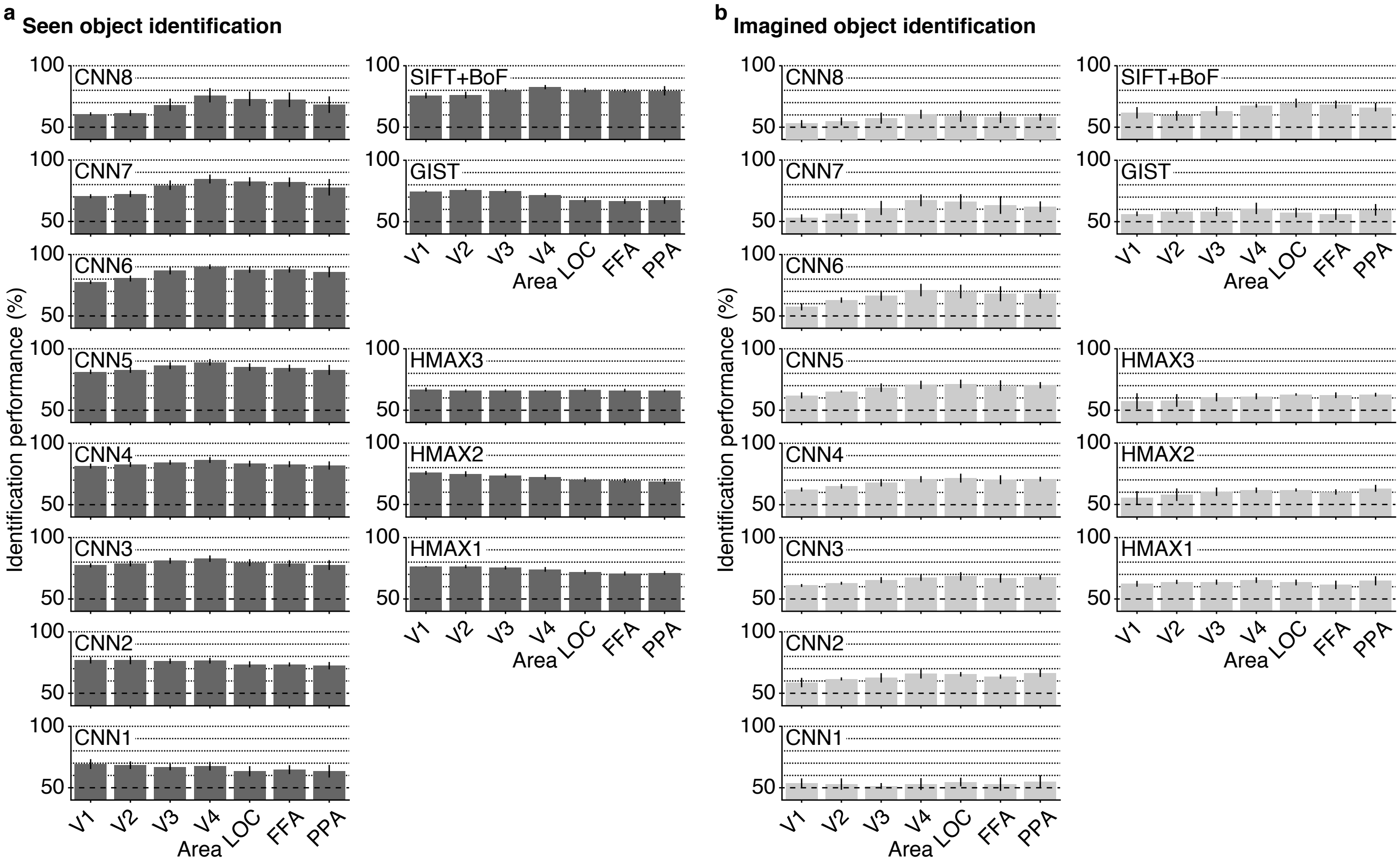

**Supplementary Figure 11 | Identification accuracy for all combinations of feature types/layers and ROIs obtained by image feature decoders.** Identification was performed for all combinations of one of the 50 test object categories and one of the 15,322 candidate categories (identification from two categories; error bars, 95% CI across five subjects; dashed line, chance level, 50%). (**a**) Seen object identification. (**b**) Imagined object identification. Both seen and imagined objects were successfully identified with most of the feature–ROI combinations (91 and 84 out of a total of 91 feature–ROI pairs for seen and imagined conditions, respectively; one-sided *t* test, uncorrected $P < 0.05$). In seen object identification, the accuracy for higher-order features tended to be better with higher ROIs, while that for lower-order features tended to be better with lower ROIs, as observed in the image feature decoding accuracy (Fig. 3b). In imagined object identification, in contrast, all feature types/layers showed a similar trend of flat or slightly elevated accuracies in higher ROIs.

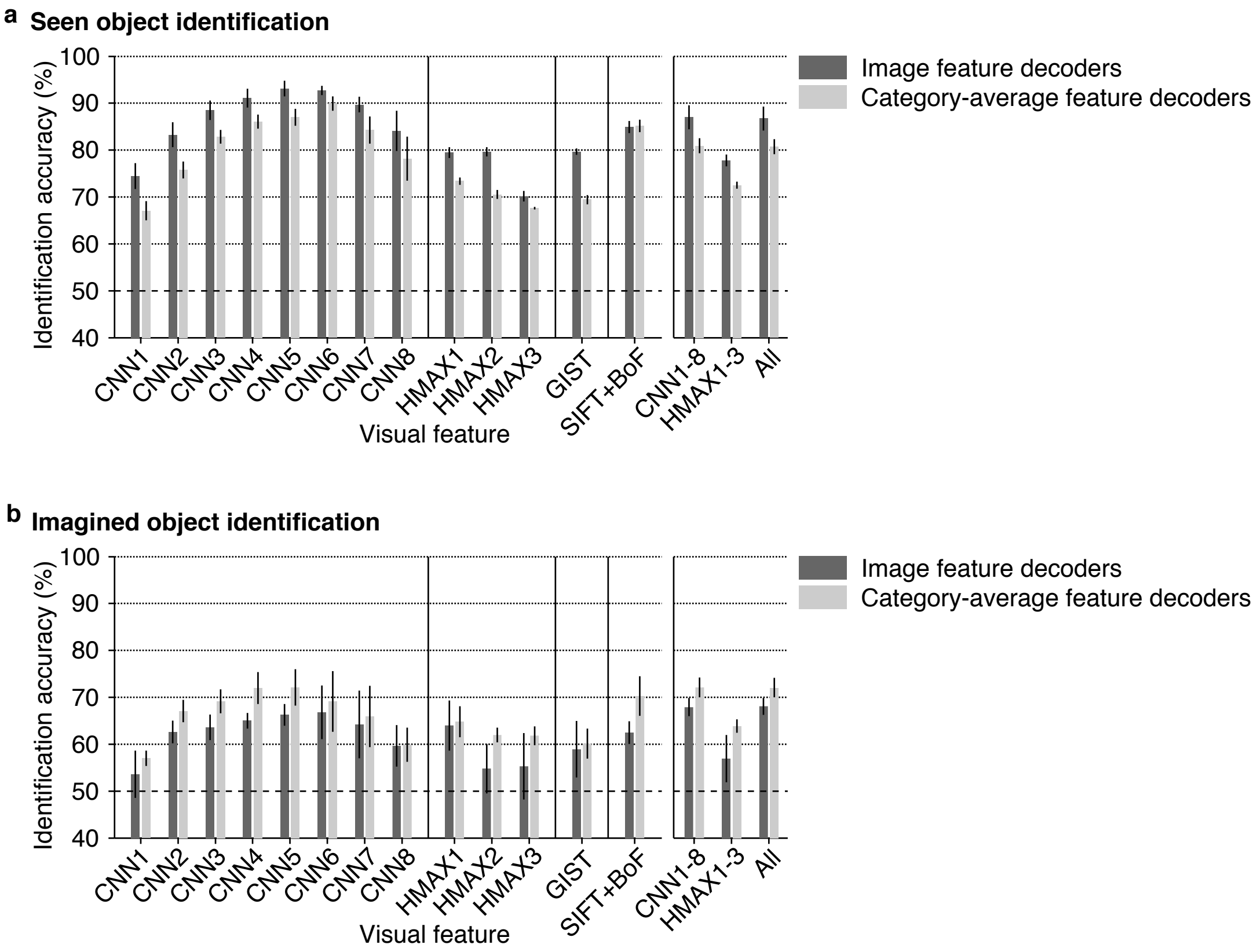

**Supplementary Figure 12 | Identification accuracy by image feature decoders and category-average feature decoders.** The same identification analyses as shown in Fig. 10a and b (image feature decoders) were performed with the decoders trained to predict category-average features of presented images (category-average feature decoders; identification from two categories; error bars, 95% CI across five subjects; dashed line, chance level, 50%). (**a**) Seen object identification. (**b**) Imagined object identification. A similar pattern of accuracy across ROIs was observed from the two types of decoders. The overall accuracy for seen object identification tended to be higher with image feature decoders than with category-average feature decoders, while that for imagined object identification tended to be lower with image feature decoders than with category-average feature decoders.

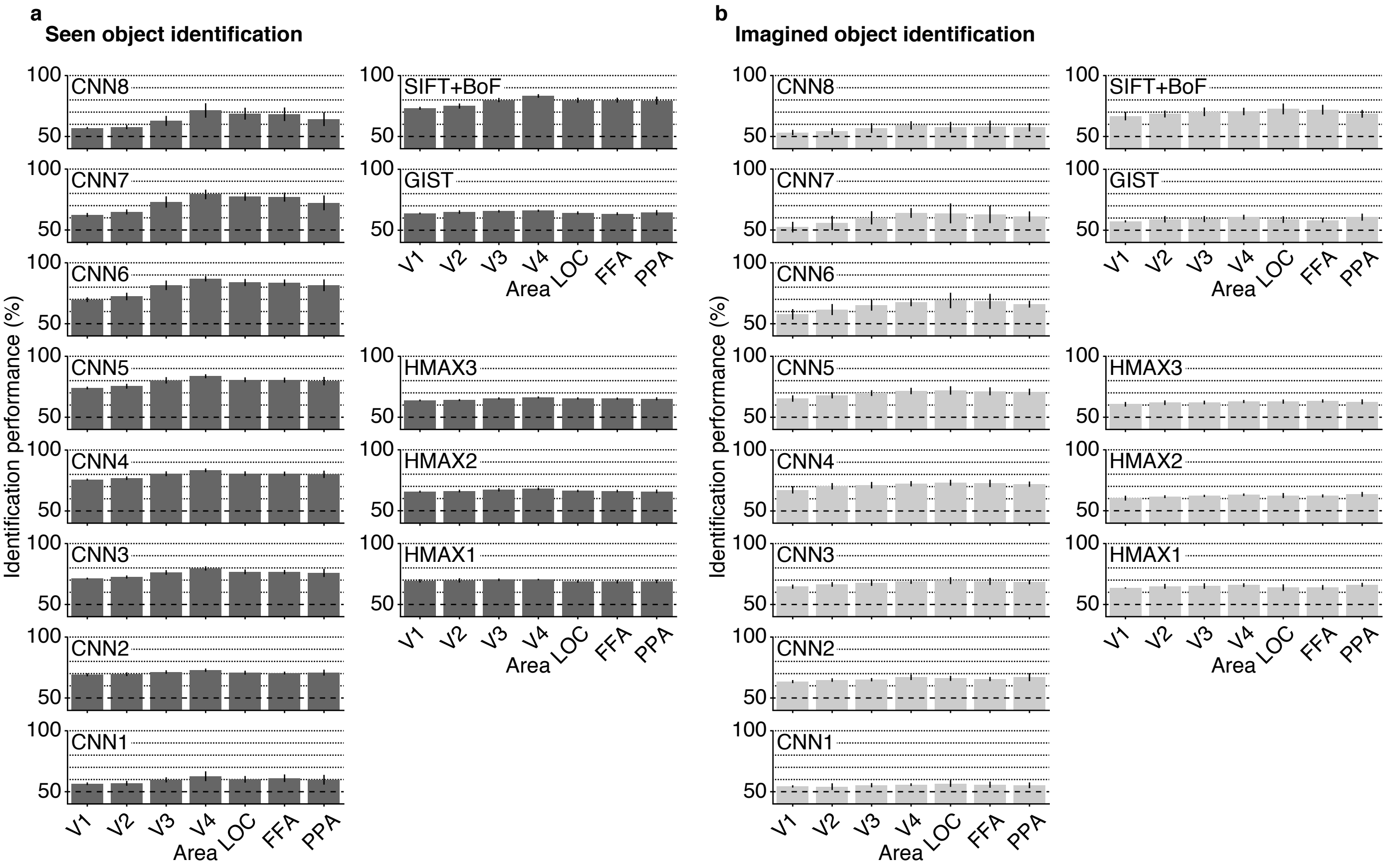

**Supplementary Figure 13 | Identification accuracy for all combinations of feature types/layers and ROIs obtained by category-average feature decoders.** The same identification analysis as shown in Supplementary Fig. 11 was performed with the decoders trained to predict category-average features of the presented images (error bars, 95% CI across five subjects; dashed line, chance level, 50%). (**a**) Seen object identification accuracy. (**b**) Imagined object identification accuracy. Both seen and imagined objects were successfully identified at a statistically significant level with most of the feature–ROI combinations (91 and 90 out of a total of 91 feature–ROI pairs for seen and imagined conditions, respectively; one-sided *t* test, uncorrected $P < 0.05$).

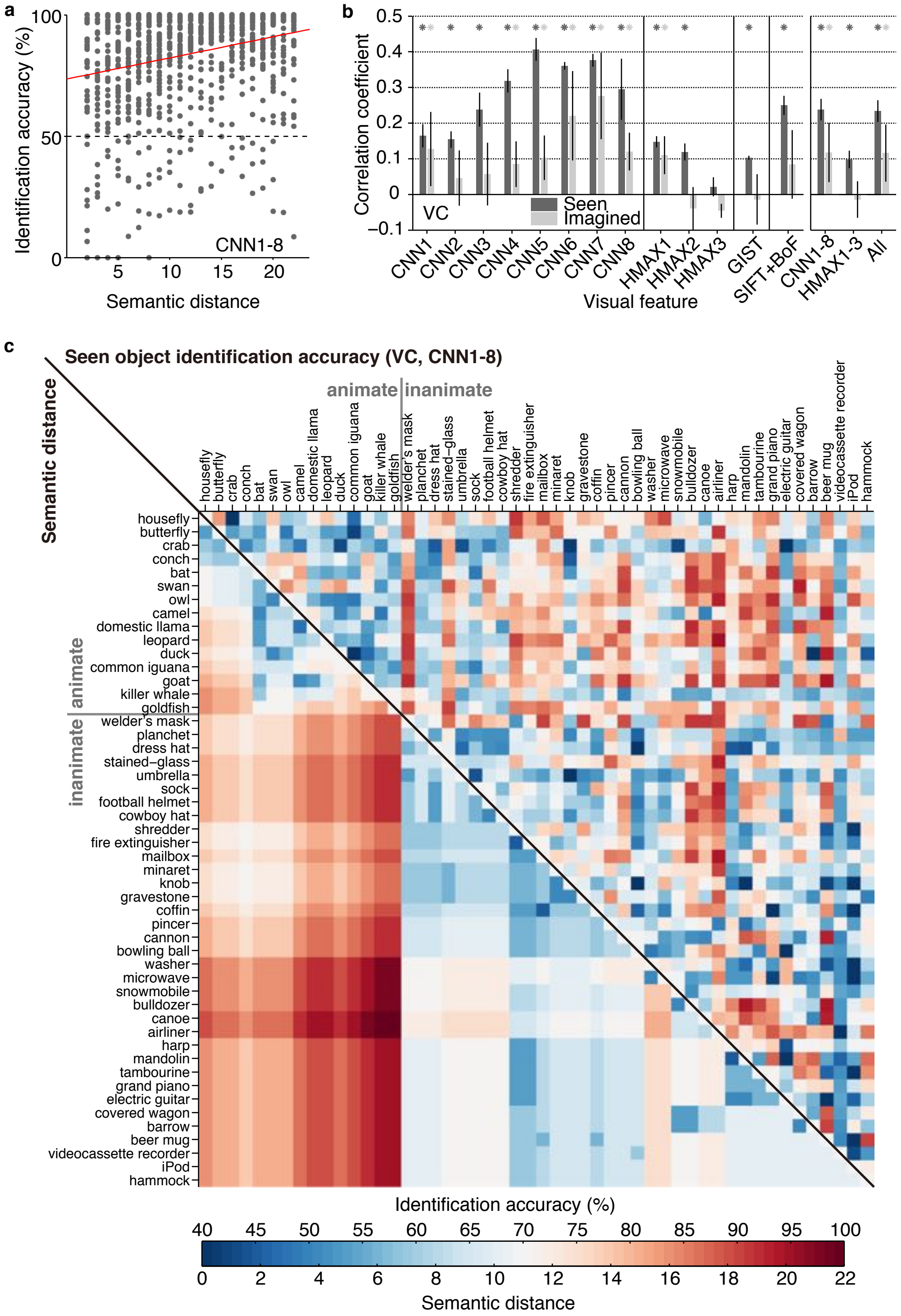
a
Identification accuracy (%)
100
50
0
5
10
15
20
CNN1-8
Semantic distance
b
Correlation coefficient
0.5
0.4
0.3
0.2
0.1
0
−0.1
VC
Seen
Imagined
CNN1
CNN2
CNN3
CNN4
CNN5
CNN6
CNN7
CNN8
HMAX1
HMAX2
HMAX3
GIST
SIFT+BoF
CNN1-8
HMAX1-3
All
Visual feature
c
Seen object identification accuracy (VC, CNN1-8)
Semantic distance
animate
inanimate
housefly
butterfly
crab
conch
bat
swan
owl
camel
domestic llama
leopard
duck
common iguana
goat
killer whale
goldfish
welder's mask
planchet
dress hat
stained−glass
umbrella
sock
football helmet
cowboy hat
shredder
fire extinguisher
mailbox
minaret
knob
gravestone
coffin
pincer
cannon
bowling ball
washer
microwave
snowmobile
bulldozer
canoe
airliner
harp
mandolin
tambourine
grand piano
electric guitar
covered wagon
barrow
beer mug
videocassette recorder
iPod
hammock
Identification accuracy (%)
40 45 50 55 60 65 70 75 80 85 90 95 100
0 2 4 6 8 10 12 14 16 18 20 22
Semantic distance

**Supplementary Figure 14 | Relation between semantic distance and identification accuracy.** Instead of evaluating mass identification accuracy by aggregating accuracies for all combinations of 50 test and 15,322 candidate categories (cf., Fig. 10a and b), identification accuracy was evaluated for each test category with candidate categories at a specified semantic distance to the test category (predicted from VC; five subjects averaged) for (**a**) and (**b**). (**a**) Semantic distance versus seen identification accuracy from concatenated vectors of CNN1–8. Each dot in the scatterplot denotes the mean identification accuracy obtained by averaging identification accuracies for all combinations of one test category and candidate categories at a specified semantic distance to the test category. The solid red line indicates a fitted regression line. (**b**) Correlation coefficients between the semantic distance and the mean identification accuracy (asterisks, one-sided *t* test after Fisher's Z transform, uncorrected $P < 0.05$). The identification accuracy and semantic distance tended to be positively correlated with each other especially with high correlation coefficients for the mid-to-high level CNN layers (CNN4–8) under both of the seen and imagined conditions. (**c**) A matrix of semantic distance and seen object identification accuracy among the 50 test categories. The semantic distance (lower triangle) and the seen object identification accuracy (upper triangle; CNN1–8; predicted from VC; five subjects averaged) are shown for all pairs of the 50 test categories. Identification accuracies (upper triangle) were calculated with fMRI data of individual trials (without averaging multiple trials corresponding to the same category) so that the accuracy with each pair can be evaluated with many instances of identification. The matrix shows a moderate level of symmetry (with respect to the diagonal line), indicating a positive correlation between the semantic distance and the identification accuracy across the pairs. The segregation between animate vs. inanimate categories[1-4] can be observed in the identification accuracy as well as in the semantic distance.

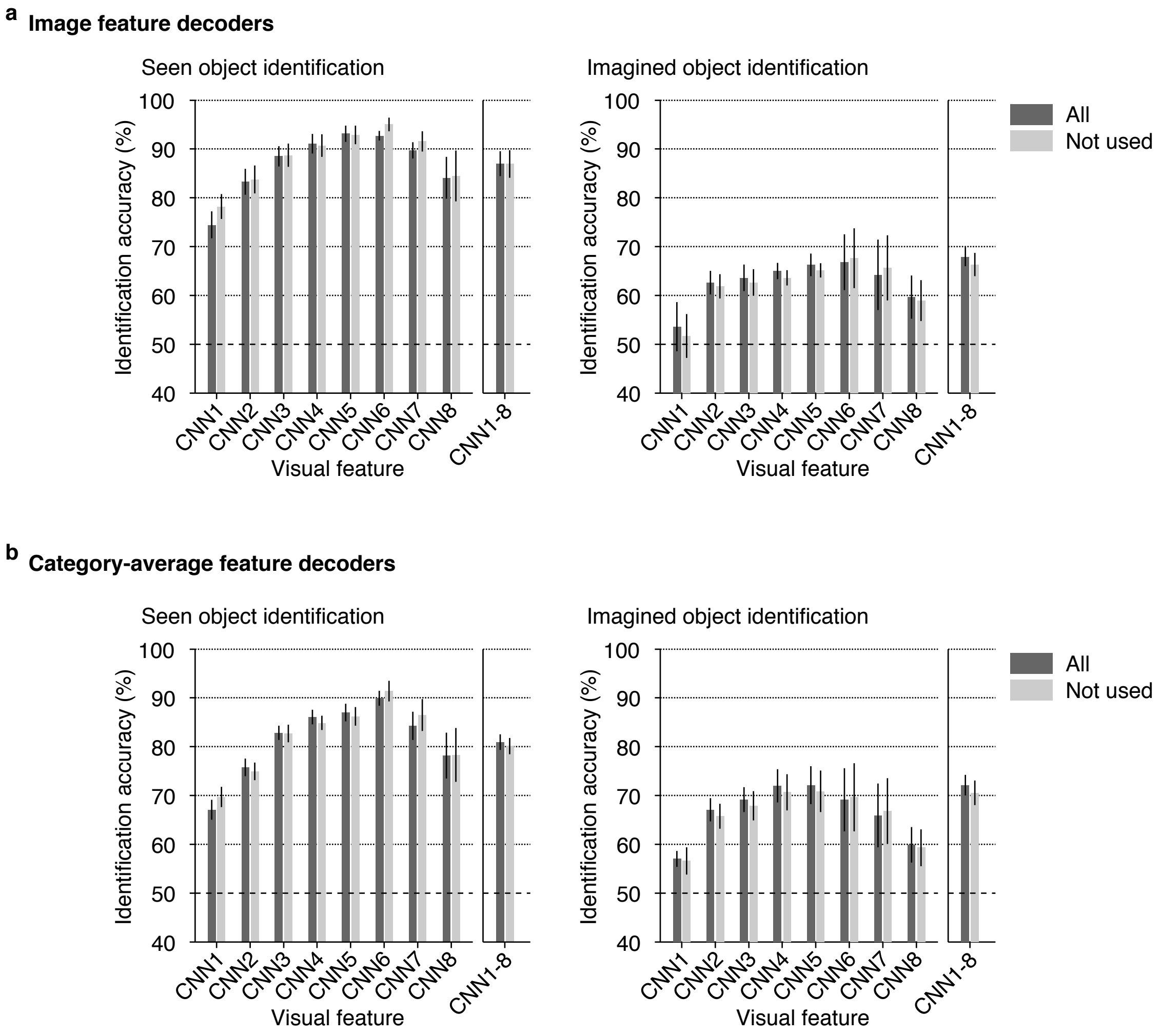

**Supplementary Figure 15 | Identification accuracy for object categories not used for CNN model training.** Mean identification accuracy for categories not used for CNN model training ($n$ = 30) were evaluated and are shown with those for all 50 test categories (identification from two categories; predicted from VC; error bars, 95% CI across five subjects; dashed lines, chance level, 50%). (**a**) Identification accuracy obtained by image feature decoders. (**b**) Identification accuracy obtained by category-average feature decoders. Identification accuracies for categories not used for CNN model training were qualitatively consistent to those for all test categories under all conditions.

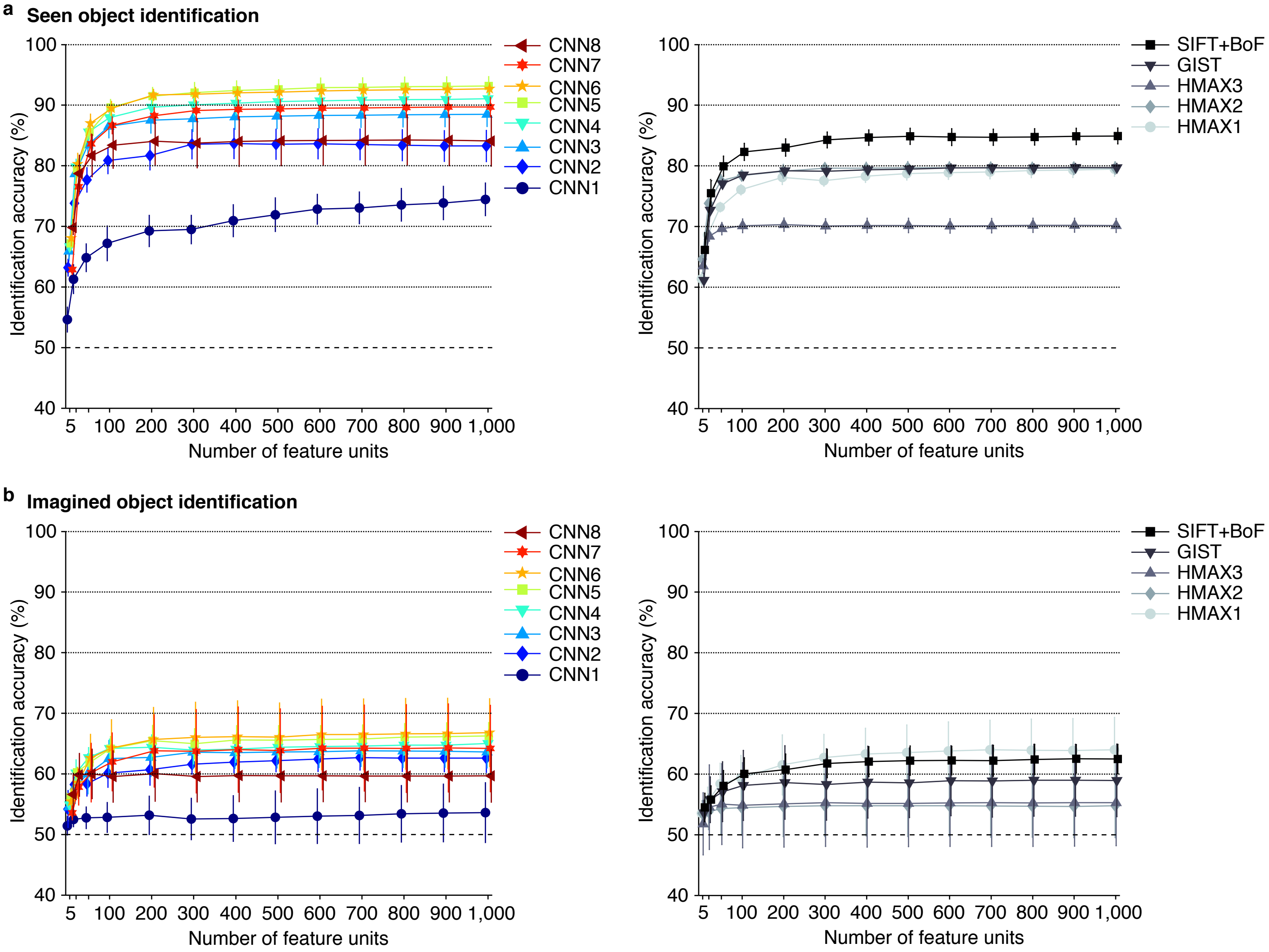

**Supplementary Figure 16 | Identification accuracy as a function of the number of feature units.** Identification was performed using a different number of feature units from each visual feature type/layer for all combinations of the 50 test object categories and 15,322 candidate categories (identification from two categories; predicted from VC by image feature decoders). The analysis was repeated 10 times for each number of feature units, and the accuracy was pooled across 10 repetitions of category candidate selection and 50 test samples (error bars, 95% CI across five subjects; dashed lines, chance level, 50%). (**a**) Seen object identification. (**b**) Imagined object identification. The accuracies of most visual features were saturated at a few hundred units. The accuracy trend across feature types/layers remained nearly constant across the number of feature units.

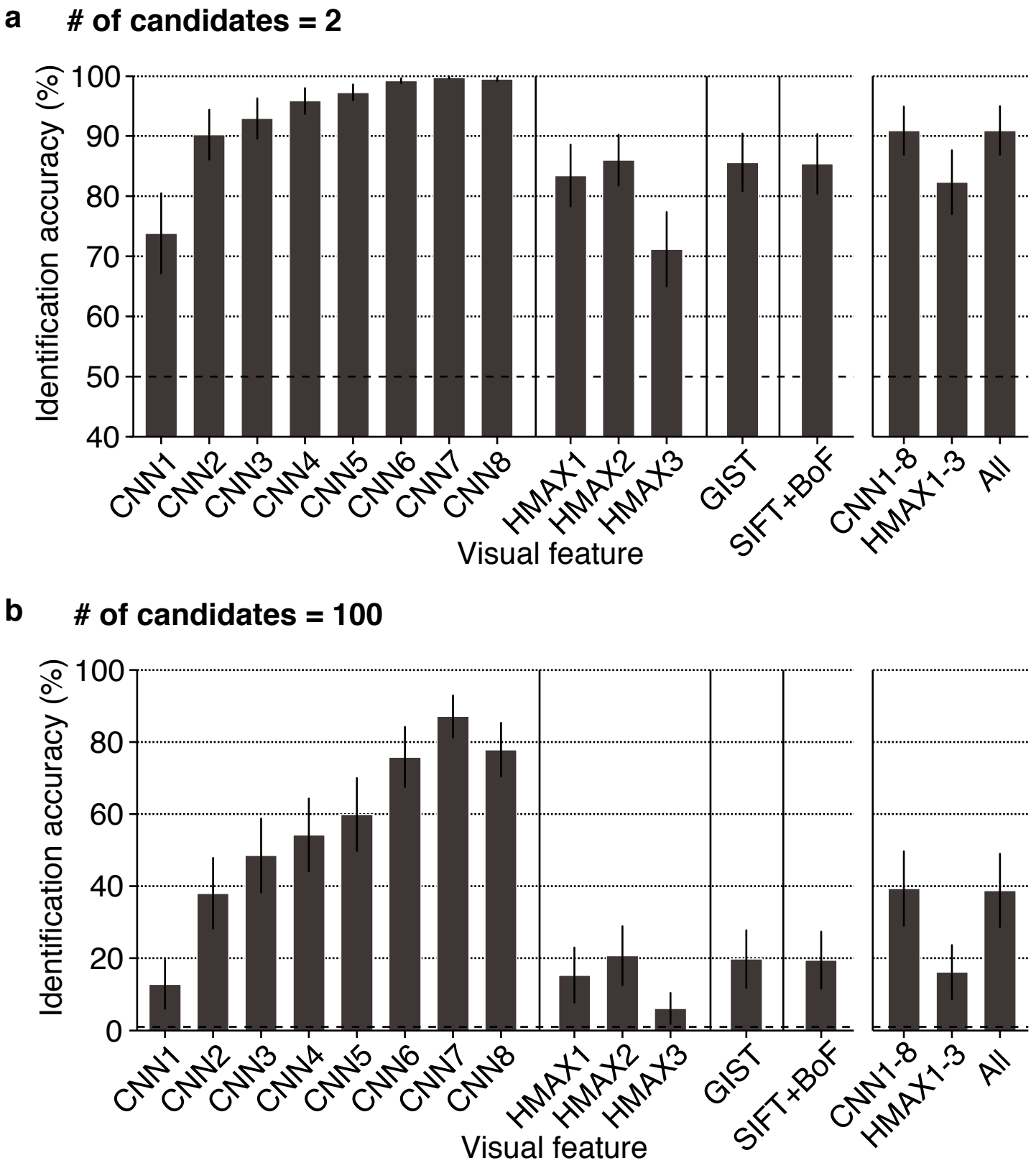

**Supplementary Figure 17 | Identification accuracy with true image feature values (generic object recognition, GOR).** The GOR identification accuracy for each visual feature type/layer is shown. The GOR accuracy is equivalent to the case when image features are perfectly predicted from brain activity using image feature decoders. (**a**) Identification from two categories. Identification was performed for all combinations of one of the 50 test object categories and one of the 15,322 candidate categories (error bars, 95% CI across 50 test categories; dashed line, chance level, 50%). (**b**) Identification from 100 categories. Identification was repeated for 100 candidate sets of randomly selected 100 categories for each of the 50 test categories. The percentage of correct identifications was averaged across the candidate sets (error bars, 95% CI across 50 test categories; dashed line, chance level, 1%). The analysis showed a slightly poorer identification with CNN8 than with CNN7. The high accuracy of original CNN features in the object recognition task would be one reason of the high accuracy of the CNN features in our generic decoding approach. While the reason why the CNN performed best among the other visual features has been debated in the field of computer vision, acquiring natural feature representations, which was shown as preferred images in Figure 4 and Supplementary Figure 4, may explain such high accuracy in object recognition.

# Supplementary References